\documentclass[runningheads]{svmult}

\usepackage{makeidx}   
\usepackage{graphicx}  
\usepackage{subeqnar}  
\usepackage{multicol}  
\usepackage{physprbb}  
\makeindex             

\begin{document}
\title*{Helioseismology: a fantastic tool to probe the interior of the Sun}
\toctitle{Helioseismology: a fantastic tool to probe the interior of the
Sun}
\titlerunning{Helioseismology: a fantastic tool to probe the interior of the Sun}
\author{Maria Pia Di Mauro}
\authorrunning{Maria Pia Di Mauro}
\institute{Teoretisk Astrofysik Center, Bygn. 520, Ny Munkegade, DK 8000 
Aarhus C, Denmark}

\maketitle 

\begin{abstract}
Helioseismology, the study of global solar oscillations,
 has proved to be an extremely powerful tool for the investigation of
the internal structure and dynamics of the Sun. 

Studies of time changes in frequency observations
of solar oscillations from helioseismology experiments 
on Earth and in space have shown, for example,
that the Sun's shape varies 
over solar cycle timescales.

In particular, far-reaching inferences about the Sun have been obtained 
by applying inversion techniques to observations of frequencies of 
oscillations. 
 The results, so far, have shown that the solar structure is remarkably close to the predictions of the standard solar model and, recently, that 
the near-surface region can be probed with sufficiently high spatial resolution as to 
allow investigations of the
equation of state and of the solar envelope helium abundance.

The same helioseismic inversion methods can be applied to the
rotational frequency splittings to deduce with high accuracy
the internal rotation velocity of the Sun, 
 as function of radius and latitude. This
also allows us to study
some global astrophysical properties of the Sun, 
such as the angular momentum, the grativational quadrupole moment and
the effect of distortion induced on the surface (oblateness).

The helioseismic approach 
 and what we have learnt from it during the last decades
 about the interior of the Sun are reviewed here.
\end{abstract}

\section{Introduction}

In the early 60's accurate observations of the photospheric spectrum revealed
the existence of  
oscillatory motions, with periods around 5 minutes, on the Sun's
surface \cite{le62}, \cite{no63}.
The observed oscillatory character of the surface was 
theoretically explained by
Ulrich \cite{ul70} and independently by Leibacher \& Stein \cite{le71}
as due to acoustic waves (i.e. $p$-modes) -- generated for some not well known reason in the convection zone and maintained by pressure force -- trapped
in resonant cavities between the Sun's surface and an inner turning point,
whose depth depends on the local speed of sound and frequency.
Only few years later
more accurate observations carried out by Deubner \cite{de75} were able to 
confirm the previous theoretical hypothesis about the modal
 nature of solar oscillations. 

The unprecedented discovery of existence of such phenomenon
 opened for the first time human eyes to the knowledge of the solar interior
and formed the basis for the development of {\it helioseismology}.
Like the geoseismology, which studies the Earth's interior through the waves produced during the earthquakes, helioseismology study the interior of the Sun 
through the small oscillations detected at the surface.

In fact, since each wave, characterized by a
specific frequency and wave number, propagates through a different region of
the Sun, probing the physical properties of the crossed medium, like
temperature and composition, it is possible to deduce the internal
stratification and dynamics of the Sun from the spectrum of resonant modes.

Since the first observations, many thousands of modes of oscillation 
have already been identified with great accuracy. The spectrum 
 extends from $0.6\, m$Hz to $5.5\,m$Hz and it has
 a maximum amplitude of about $15\,cm/s$ in velocity.
This incredible amount of information collected contributed to the success 
of this discipline and
has permitted a deep knowledge of the Sun, not imaginable thirty years ago.

\section{Theoretical approach to helioseismology}

\subsection{Basic equations of adiabatic oscillations}

A star, like the Sun, is a gaseous sphere in hydrostatic equilibrium,
and the oscillations are fluid-dynamical phenomena caused
 by the action of a restoring force which arises when the original
 equilibrium status is perturbed.
Such hydrodynamical systems can be described by specifying 
all the physical quantities as functions of 
the position $\vec{r}$ and time $t$.

To provide a background for the treatment of stellar pulsations, 
it is useful to consider briefly
the basic equations of hydrodynamics, which
can be derived by applying the
fundamental principles of conservation of mass, of momentum and of energy.
We neglect viscosity and magnetic fields and
 we can also assume that gravity is the only acting body force and 
that the radiation is the dominant contribution to the flux of energy.

The conservation of mass is expressed by
the equation of continuity:
\begin{equation}
\frac{\D \varrho}{\D t}= -\varrho ~\  \mathrm{div}\,\vec{v}\; ,
\label{eq1}
\end{equation}
where $\varrho=\varrho(\vec{r},t)$ is the density and
$\vec{v}=\D \vec{r}/\D t$ is the local velocity.
The equation of motion is:
\begin{equation}
\varrho \frac{\D \vec{v}}{\D t} = -\nabla p +\varrho \nabla 
\vec{\mathrm{\Phi}}\; ,
\end{equation} 
where $p=p(\vec{r},t)$ is the pressure and $\vec{\mathrm{\Phi}}$ is the gravitational 
potential which satisfies the
Poisson's Equation, 
\begin{equation}
\nabla ^{2} \vec{\mathrm{\Phi}} = -4 \pi G \varrho\; ,
\end{equation}
$G$ being the gravitational constant.
And finally the energy equation:
\begin{equation}
\frac{\D Q}{\D t}=\frac{\D E}{\D t}+p\frac{\D}{\D t}\left(\frac{1}{\varrho}\right)=
\varepsilon-\frac{\mathrm{div} \vec{F}}{\varrho}\; ,
\label{eq4}
\end{equation}
where ${\D Q}/{\D t}$ is the rate of heat loss or gain per unit of mass, $E$ is the internal energy in unit of mass, $\varepsilon$ is the rate of energy generation per unit mass,
while $\vec{F}$ is the flux of energy.

By using thermodynamic identities the energy equation can be expressed in terms of other, more convenient variables, like in \cite{co68}:
\begin{equation}
\frac{\D Q}{\D t}=\frac{1}{\varrho(\Gamma_3 -1)}\left(\frac{\D p}{\D t}-\frac
{\Gamma_1 p}{\varrho}\frac{\D \varrho}{\D t}\right)\; ,
\label{ad}
\end{equation}
where $\Gamma_1$ and $\Gamma_3$ are the first and the third adiabatic exponent,
defined by:
\begin{equation}
\Gamma_1=\left(\frac{\partial \ln p}{\partial \ln \varrho}\right)_{ad}\,\,,
\Gamma_3-1=\left(\frac{\partial \ln T}{\partial \ln \varrho}\right)_{ad}\; ,
\label{gamma}
\end{equation}
where $T$ is the temperature and the derivatives are
calculated at constant specific entropy. 

The observed amplitude of solar oscillations are very small $(\delta r/R_{\odot}\simeq 10^{-4})$, so that the pulsations can be described with accuracy by
applying a linear perturbation analysis of 
the basic equations of hydrodynamic.

Let us consider a static equilibrium model
with pressure $p_0(r)$, density $\varrho_0(r)$ etc.
If we consider {\it Eulerian} perturbations, 
the generic physical quantity $f$ 
 can be written in the following way:
\begin{equation}
f(\vec{r}, t)=f_{0}(\vec{r})+ f'(\vec{r}, t)\; ,
\end{equation}
where $f_{0}(\vec{r})$ is the unperturbed term and
 $f'(\vec{r}, t)$ is the small perturbation at a given spatial point.
The small perturbation can also be written in the Lagrangian form, by considering a frame following the motion of
an element of gas which moves from position $\vec{r}$ to $\vec{r}+\vec{\delta r}$:
\begin{equation}
\delta f(\vec{r})=f(\vec{r}+\vec{\delta r})-f_{0}(\vec{r})=f'(\vec{r})+\vec 
{\delta r} \cdot \nabla f_{0}\;. 
\end{equation}

Since the typical pulsation period is much smaller than the time required to dissipate the thermal energy, we first assume that
the adiabatic approximation is sufficient to discuss the dynamical characteristics in the interior of the Sun.
This hyphotesis, which greatly simplifies the treatment of stellar pulsations, is not longer verified in the very superficial layers and nonadiabatic effects on the frequencies should not be neglected in the study of the surface 
layers and of the pulsation energetics.

According to these assumptions and by perturbing the four basic equations
(\ref{eq1})--(\ref{eq4}), we obtain the following system of four linear 
equations in four unknowns to study the small oscillations under adiabatic conditions:
\begin{subeqnarray}
\varrho' &=& -\;\mathrm{div}\;(
\varrho_{0}\vec{\delta r}) \; \label{eqosc1},\\     
\varrho_{0} \frac{\partial\vec{v}}{\partial t }&=& 
-\nabla p'+\varrho'\nabla \vec{\mathrm{\Phi}} 
+\varrho_{0} \nabla \vec{\mathrm{\Phi}'}  \; ,\\
\nabla^{2} \vec{\mathrm{\Phi'}} & = &-4 \pi G \varrho' \; , \\
p'+\vec{\delta r}\cdot \nabla p_{0}&=&
\frac{\Gamma_{1}p_{0}}{\varrho_{0}}\left(\varrho'+\vec{\delta r} \cdot 
\nabla \varrho_{0}\right)\;.
\label{eqosc2}
\end{subeqnarray}
An extensive and detailed derivation of the equations of adiabatic oscillations
is provided by Unno et al. in \cite{un89}, by Christensen--Dalsgaard \& Berthomieu in \cite{CD91} or by
Christensen--Dalsgaard in \cite{CD98}.

\subsection{Spherical Harmonic Representation}
The general equations for small oscillations presented above must now be
 derived in the specific case of stars, which are assumed to have
 a spherically symmetric and a time-independent equilibrium structure.
\begin{figure}[bt]
\begin{center}
\includegraphics[width=.6\textwidth]{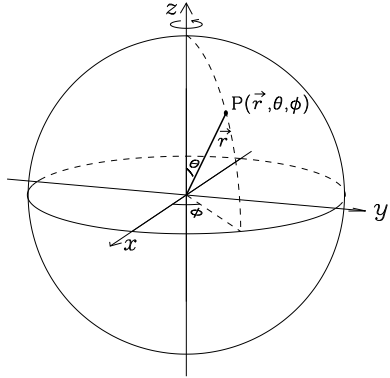}
\end{center}
\caption[]{The spherical polar coordinate system \label{1}}
\end{figure}

Let consider a spherical polar coordinates  system ($r,\theta,\phi$), where $r$ is the distance to the centre, $\theta$ is the colatitude, and $\phi$ is the longitude (Fig. \ref{1}).
In the polar coordinates system a vector 
field can be written by specifying its components in the 
radial and angular directions.
Thus the displacement, for example, can be written as:
\begin{equation}
\vec{\delta r}(r,\theta,\phi,t)=\xi_r \vec{a}_r+\xi_\theta\vec{a}_\theta+ \xi_{\phi}
\vec{a}_{\phi}\; ,
\end{equation}
where $\vec{a}_r$, $\vec{a}_{\theta}$ and $\vec{a}_{\phi}$ are 
the unit vectors in 
the $r$, $\theta$ and $\phi$ directions respectively.
By introducing $\xi_h$, the horizontal
 component of the vector, we can also write:
\begin{equation}
\vec{\delta r}=\xi_r \vec{a}_r+\xi_h\vec{a}_h\; ,
\end{equation}
where  $\xi_{r}=\xi_{r}(r,\theta,\phi,t)$ and $\xi_{h}=\xi_h(r,\theta,\phi,t)$ are the radial and horizontal components
of the displacement.

Since the 
equilibrium state depends only on the radius $r$,
the solutions of the linear system (\ref{eqosc1})--(\ref{eqosc2}) 
can be obtained in the following form, by separating the spatial 
from the temporal dependence:
\begin{equation}
f'(r,\theta,\phi,t)=
\tilde{f}'(r)f(\theta,\phi)\exp({-i\omega t})\; ,
\label{pert}
\end{equation}
where the time dependence has been expressed in terms of an harmonic function,  characterized by a frequency $\omega$, the amplitude $\tilde{f}'(r)$ is a function of $r$ alone, and $f(\theta,\phi)$ describes the angular variation of the solution.

In the spherical symmetric system, all derivatives with respect to $\theta$ and $\phi$ can be expressed in the form of the tangential Laplace operator:
\begin{equation}
\nabla^{2}_{h}\equiv \frac{1}{r^{2} \sin \theta}\frac{\partial}{\partial \theta}
\left( \sin \theta \frac{\partial}{\partial \theta} \right)+\frac{1}{r^{2} \sin 
^{2}\theta} \frac{\partial^{2}}{\partial ^{2} \phi}\; .
\end{equation}
Consequently, $f(\theta,\phi)$
can be found as eigenfunction of $\nabla^{2}_{h}$
and it may be chosen to be
the spherical harmonic $Y_{l}^{m}(\theta, \phi)$  
of degree $l$ and azimuthal order $m$,
 which indeed satisfies the eigenvalues problem:
\begin{equation}
\nabla^{2}_{h}Y^{m}_{l}(\theta, \phi)=-\frac{l(l+1)}{r^2}Y_{l}^{m}
(\theta,\phi)=-k^2_hY_{l}^{m}
(\theta,\phi)\; ,
\label{kh}
\end{equation}
where $l$ and $m$ are integers, such that $-l\leq m\leq l$, and $k_h$ is the horizontal component of the wave number.
The spherical harmonics $Y_{l}^{m}(\theta,\phi)$
are defined by:
\begin{equation}
Y_{l}^{m}(\theta, \phi)=N_{m,l}P_{l}^{m}(\cos\theta)e^{im\phi}\; ,
\end{equation}
where $P_{l}^{m}$ is the associated Legendre polynomial and $N_{m,l}$ is a constant such that the following integral over the unit sphere is satisfied:
\begin{equation}
\int_{0}^{2 \pi}\int_{0}^{\pi} Y_{l}^{m}(\theta, \phi) Y_{l'}^{m'}
(\theta, \phi)
\sin \theta d \theta d \phi=\delta_{ll'} \delta_{mm'}\; ,
\end{equation}
where $\delta_{ll'}$ and $\delta_{mm'}$ are Kronecker's deltas, so that the integral is zero if $l\neq l'$ and $m\neq m'$.

It follows that the perturbation quantities (\ref{pert}) can be written as:
\begin{equation}
f'(\vec{r}, \theta, \phi, t)=\sqrt{4\pi}\tilde{f}'(r)Y_{l}^{m}(\theta, \phi) e^{-i \omega t}
\end{equation}
and that the displacement vector can be expressed by:
\begin{equation}
\vec{\delta r}=\sqrt{4\pi}\Re\left\{\left[{\tilde{\xi}_{r}}(r)
\vec{a}_{r}+
\tilde{\xi}_{h}(r)\left(\frac{\partial}{\partial \theta}\vec{a}_{\theta}
+\frac{\partial}{\sin \theta \, \partial
\phi}\vec{a}_{\phi}\right) \right]Y_{l}^{m}(\theta, \phi)e^{-i \omega t}\right\}\; ,
\end{equation}
where $\Re$ stands for the real part.

By substituting the spherical harmonic representations into
 Eqs. (\ref{eqosc1})--(\ref{eqosc2}),
we obtain the following set of ordinary differential equations, which 
describes
the stellar adiabatic oscillations:
\begin{subeqnarray}
\frac{\D \xi_{r} }{\D r}=-\left(\frac{2}{r}+\frac{1}{\Gamma_{1} p}
\frac{\D p}{\D r}\right)\xi_{r}+
\frac{1}{\varrho c^2} \left( \frac{S^2_l}{\omega^{2}}-1\right)p'-\frac{l(l+1)}{r^{2}\omega^{2}} \vec{\mathrm{\Phi}'} \;, \label{eqmod1} \\
\frac{\D p'}{\D r}=\varrho(\omega^{2}-
N^{2})\xi_{r}+\frac{1}{\Gamma_1 p}\frac{\D p}{\D r}p'+\varrho
\frac{\D \vec{\mathrm{\Phi}'}}{\D r}\; , \\
\frac{1}{r^{2}}\frac{\D}{\D r}\left( r^{2} \frac{\D \vec{\mathrm{\Phi'}}}{\D r} \right)=
-4\pi G\left(\frac{p'}{c^2}+\frac{\varrho\xi_r}{g}N^{2}\right)
+ \frac{l(l+1)}{r^2} \vec{\mathrm{\Phi}'}\; ,
\label{eqmod3}
\end{subeqnarray}
where
$S_l$ is the Lamb frequency
\begin{equation}
S_l^2=\frac{l(l+1)c^2}{r^2}\; ,
\end{equation}
$N$ is the buoyancy frequency
\begin{equation}
N^2=g\left(\frac{1}{\Gamma_1 p}\frac{\D p}{\D r}-\frac{1}{\varrho}
\frac{\D \varrho}{\D r}\right)\; ,
\end{equation}
and $c$ is the speed of the sound in adiabatic conditions, such that
under the reasonable assumption that the stellar interior
can be approximate to an ideal gas:
\begin{equation}
c^2=\frac{\Gamma_1 p}{\varrho}\simeq\frac{\Gamma_1 k_B T}{\mu m_u}\; ,
\label{c2}
\end{equation}
where $k_B$ is the Boltzmann's constant, $\mu$ is the mean molecular weight and $m_u$ is the atomic mass unit.

\begin{figure}[bt]
\hspace{0.8cm}
\includegraphics[width=.9\textwidth]{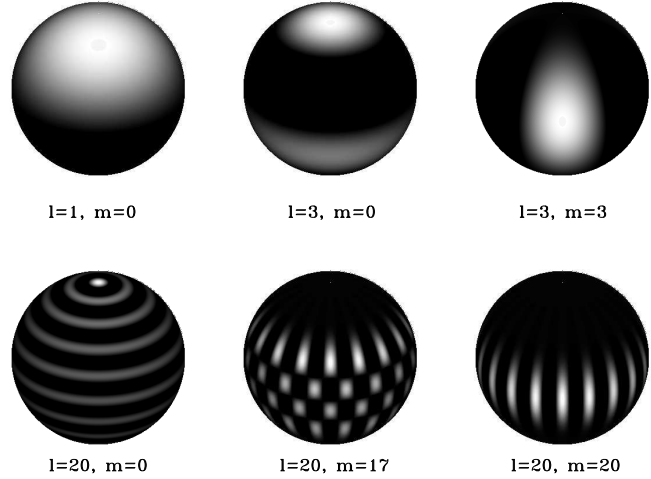}
\caption[]{Illustration of some spherical harmonics. For clarity the polar 
axis has been rotated of $30^{\circ}$ into respect the plane of the page. 
Positive patterns are indicated by brighter surfaces while negative patterns 
are darker \label{Ylm}}
\end{figure}

The fourth-order system of ordinary differential equations
(\ref{eqmod1})--(\ref{eqmod3}), together with appropriate boundary conditions
at centre $r=0$ and at the surface $r=R_{\odot}$,
 constitutes an eigenvalue problem, which admits
solutions only for particular values of the eigenfrequencies $\omega$.
The discrete set of solutions, obtained for each $(l,\, m)$
 is labelled with an integer $n$ and describes the so-called
{\it spheroidal modes}.
The modes are therefore identified by three quantum numbers, the radial order
 $n$, which is the number of the nodes of the wave in the radial direction, 
the harmonic degree $l$, which is the number of nodes on the surface in the 
direction of the latitude and
 the azimuthal order $m$, which specifies the number of the nodes along the 
longitude on the surface.
A few examples of spherical harmonics are shown in Fig. \ref{Ylm}.
It should be noticed, that in the case of spherical symmetry there are not 
preferential directions on the sphere, therefore the modes show
$(2l+1)$-fold degeneracy in $m$ and both the eigenfrequencies and the 
eigenfunctions do not depend on $m$.

\section{Propagation of solar oscillations}
 
It can be noticed that the set of equations of adiabatic oscillations 
(\ref{eqmod1})--(\ref{eqmod3}) can be 
easily solved, once the boundary conditions are known. However, an accurate 
interpretation of the results requires a more complete description 
of the phenomenon.
This
can be obtained by an asymptotic analysis of the pulsation equations, 
justified by the fact that the acoustic modes observed in the Sun,
show fairly high radial order and high degree.

The asymptotic analysis is usually carried out in the Cowling approximation
\cite{co41}, by neglecting the perturbation $\vec{\mathrm{\Phi}'}$
of the gravitational potential.
In this case the oscillations equations (\ref{eqmod1})--(\ref{eqmod3}) 
reduce to a second-order system,
that according to \cite{de84} it can be written in 
the following approximate expression:
\begin{equation}
\frac{\D ^{2}\Psi}{\D r^{2}}+\frac{1}{c^2}\left[\omega^2 -
\omega_c^{2}-S_l^{2}\left(1-\frac{N^2}{\omega^2}\right)\right]\Psi=0 \; ,
\label{eqpro}
\end{equation}
where
\begin{equation}
\Psi(r)=c^2\varrho^{1/2}\mathrm{div}\vec{\delta r}.
\end{equation}
The acoustical cut-off frequency
 $\omega_c$ is defined by:
\begin{equation}
\omega^2_c=\frac{c^2}{4H^2}\left(1-2\frac{\D H}{\D r}\right)\; ,
\end{equation}
where $H=-(\D \ln \varrho/\D r)^{-1}$ is the density scale height.

In a star in which the main body forces acting are the pressure and the 
gravity, two kind of oscillations can be maintained: the pressure or acoustic 
waves and the internal gravity waves, which form the classes of 
p modes and g modes respectively.
The dispersion relation which describes the acoustic and gravity waves 
propagation in a medium, 
 can be derived 
from Eq. (\ref{eqpro}) as:
\begin{equation}
c^2k^2_r=\omega^2-\omega_c^{2}-S_l^{2}\left(1-\frac{N^2}{\omega^2}\right)\; ,
\end{equation}
where $k_r$ is the radial component of the wave number, which clearly 
depends
on the variation of the characteristic frequencies $S_l$, $N$ and $\omega_c$ 
with radius.

\begin{figure}[t]
 \begin{center}
\includegraphics[width=.6\textwidth, angle=270]{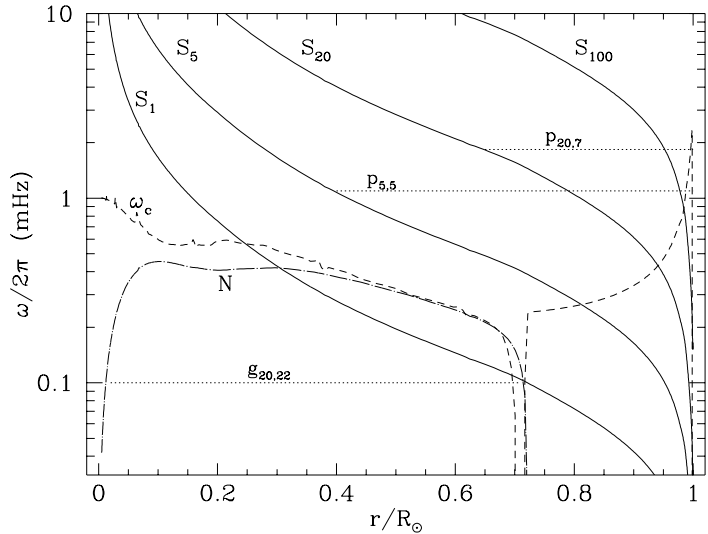}
\end{center}
\caption[]{Propagation diagram of the characteristic frequencies $N$, 
$\omega_c$ and $S_l$, calculated for some values of $l$, as functions of the fractional radius for a standard solar model. The horizontal lines indicate the trapping regions for a g mode with $l=20$ and $n=22$, and two p modes 
with ($l=5 \, ,n=5$) and ($l=20\,,n=7$)}\label{Prop}
\end{figure}

The propagation of modes of oscillation requires that $k^2_r>0$:
\begin{equation}
\omega^2 -
\omega_c^{2}-S_l^{2}\left(1-\frac{N^2}{\omega^2}\right)>0.
\label{con}
\end{equation}
It follows that the
 Eq. (\ref{con}) is satisfied
in the two domains where:
\begin{equation}
\omega^2>S_l^2\,\,\,\,\,\,\,\,\,\omega^2>\omega_c^2
\label{con1}
\end{equation}
and
\begin{equation}
\omega^2<N^2\;.
\label{con2}
\end{equation}
The conditions (\ref{con1}) and (\ref{con2}) define
 the {\it trapping regions} of p modes and g modes respectively, as illustrated in Fig. \ref{Prop}. Outside these regions
 the waves are evanescent and do not show oscillatory character in space and their amplitude decays exponentially.

\begin{figure}[t]
 \begin{center}
\includegraphics[width=.75\textwidth]{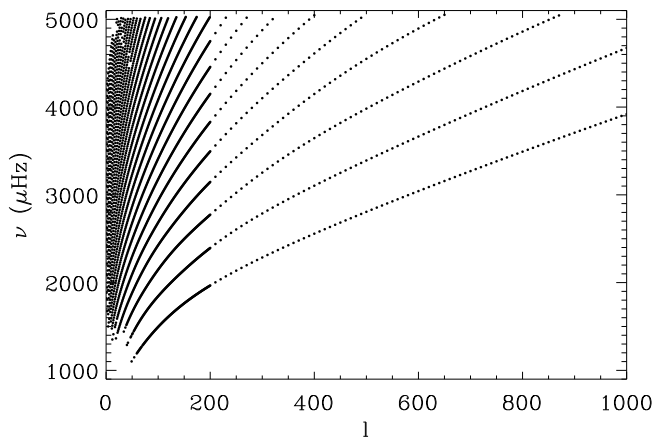}
\end{center}
\caption[]{A set of p-mode frequencies \cite{rh98}, as function of $l$,
obtained by MDI instrument on board of SOHO. Each ridge contains modes with equal values of $n$}
\label{spectrum}
\end{figure}

Detection of g modes would be extremely valuable since they have
highest amplitudes in the core, and hence their frequencies, if detected,
should be very sensitive to the structure and rotation of the deeper interior of the Sun. Unfortunately,
although claims for detection of g modes have been made \cite{ga98}, we still not have
any confirmation 
that they are really excited in the Sun, and the observed 
five-minutes oscillations correspond only to p and f modes.
Figure \ref{spectrum} shows
a set of p modes frequencies obtained in 1996 \cite{rh98}
by the MDI \cite{sc95} instrument on board the SOHO satellite.

The f modes correspond approximately to surface gravity waves
with the condition that $\mathrm{div}(\vec{\delta r})\simeq 0$, so that according to
Eqs. (\ref{eqosc1})--(\ref{eqosc2}), it is possible to assume that 
$\delta p\simeq\delta \varrho\simeq0$.
The dispersion relation of the f modes is:
\begin{equation}
\omega^2\simeq g_{\odot} k_h\; ,
\label{fmodes}
\end{equation}
where $g_{\odot}=GM/R_{\odot}^3$.
Thus, frequencies depend only on the mean density of the star, but not on 
its detailed internal structure.

\subsection{Properties of the acoustic modes}

\begin{figure}[tb]
 \begin{center} 
\includegraphics[width=.6\textwidth]{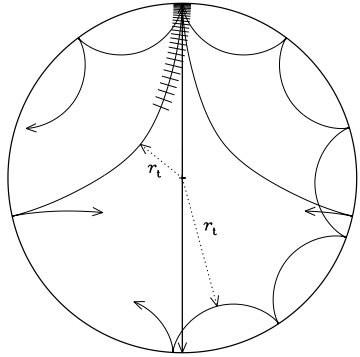}
\end{center}
\caption[]{Propagation of rays of sound in the solar interior in the case
of two p modes with degrees $l=5$ and $l=15$.
 The acoustic waves are reflected at the surface owing to the rapid decrease of density, and at the inner turning point $r_t$ due to the increase of the temperature with depth.
Notice that waves with a smaller wavelength corresponding to a higher value of the degree $l$, penetrate less deeply
\label{F.4}}
\end{figure}

The propagation of p modes in the interior of the Sun
 can be interpreted very simply in geometrical terms, by studying the 
behaviour of rays of sound, as illustrated in Fig. \ref{F.4}.
Locally the acoustic modes can be approximated by plane sound waves whose 
dispersion relation is:
\begin{equation}
\omega^2=c^2|\vec{k}|^2=c^2(k_r^2+k_h^2)\; ,
\label{om2}
\end{equation}
where $k_r$ and $k_h$ are the radial and horizontal components of the wave vector $\vec{k}$.
This means that the properties of the modes are entirely controlled by the variation of the adiabatic sound speed $c$, which depends on temperature 
(Eq.~\ref{c2}).
From Eq. (\ref{om2}), by using the definition of $k_h$ given in
 Eq. (\ref{kh}), it follows that:
\begin{equation}
k_r^2=\frac{\omega^2}{c^2}-\frac{l(l+1)}{r^2}=\frac{\omega^2}{c^2}\left(1-\frac{S_l^2}{\omega^2}\right)\; .
\label{kr}
\end{equation}
At the surface, where $c$ is small, $k_r$ is large and hence the 
wave propagates
almost vertically. Due to the increase of the sound speed with temperature, 
 $k_r$ decreases with depth, 
while $k_h$ increases as $r$ decreases,
until $k_r=0$ and the wave travels mostly horizontally.
 This condition
is reached at the turning point $r_t$, where:
\begin{equation}
\frac{c(r_t)}{r_t}=\frac{\omega}{\sqrt{l(l+1)}}\; .
\label{rt}
\end{equation}
At the turning point, the wave is gradually refracted 
and goes back towards the surface.
For $r<r_t$, $k_r$ is imaginary and the wave decays exponentially.

It appears clear from the Eq. (\ref{rt}) that lower is the harmonic degree $l$,
the deeper is located the turning point of the mode (Fig. \ref{F.4}).
Radial acoustic modes with $l=0$ penetrate to the centre, while the modes of
 highest harmonic degree observed in the Sun ($l\simeq 1000$) are trapped in 
the outer $0.2\%$ of the solar radius.

\begin{figure}[tb]
 \begin{center}
\includegraphics[width=.8\textwidth]{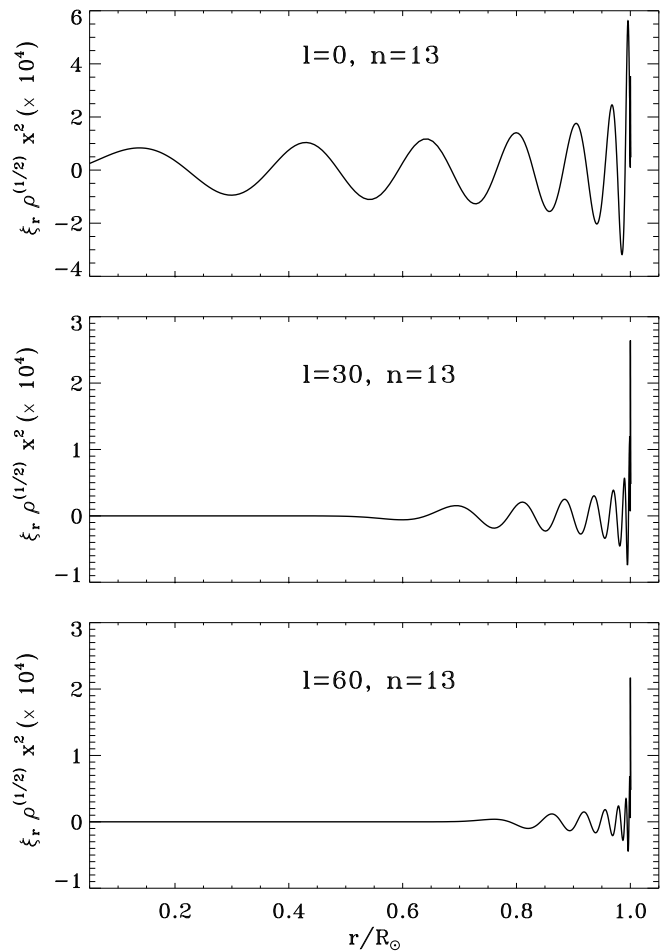}
\vspace{0.5cm}
\end{center}
\caption[]{Eigenfunction for p modes with different harmonic degree
as function of the fractional radius $x=r/R_{\odot}$. 
Here, the oscillation behaviour is enhanced, by 
scaling the eigenfunctions with the square root of the density and the squared fractional radius}\label{aut}
\end{figure}

Figure \ref{aut} shows eigenfunctions for a selection of p modes with different
degree: with increasing degree the p modes become confined closer and closer to the surface.

\section{Helioseismic investigations}

The frequencies of 
the solar oscillations depend on the structure of the equilibrium model, 
predominantly on the local adiabatic speed of sound
and in addition on the variation of the density and of the adiabatic gradient
in the Sun. 

Moreover, the oscillation frequencies have several advantages over 
all the other solar observables:
 they can be observed with great accuracy and different modes probe 
the characteristics of different layers in the interior of the Sun.
Thus, accurate observations of the acoustic frequencies, available today
 from a variety of helioseismology experiments 
on Earth and in space, 
can be used to probe the characteristics and the 
details of the interior of the Sun.

The goal of the helioseismology is, in fact, to infer the internal properties
of the Sun and to understand the physical mechanisms which govern 
the behaviour of our star.
This can be pursued by two different complementary strategies.
The first is the forward approach which consists in comparing the observed data
with the theoretical frequencies computed for a solar model,  following the analysis explained in the previous sections.
 The second is based on the use of
the observed data to deduce the internal structure and rotation of the Sun by means of data inversion.
 The inverse approach and its results will be extensively discussed
in the next sections.

It appears clear that
all the helioseismic investigations require the use of a solar model resulting from evolution of the structure equations from its formation to the present
 age. 
 The computed models depend on assumptions about the 
physical properties of matter in stars, in particular the equation of 
state, the opacity and the rates of nuclear reactions.
 
It is also necessary that the models agree with the known non-seismic properties of the Sun: the photospheric radius 
$R_{\odot}=(6.9699\pm0.07) \times 10^{10}\, cm$ \cite{al76}, 
the observed luminosity $L_{\odot}=(3.846\pm0.005)\times 10^{33} \,erg\, s^{-1}$
 \cite{wi88}, the mass $M_{\odot}=(1.989\pm 0.0004)\times 10^{33}\, g$ as obtained from the study of planetary motion, the composition of the photosphere as inferred 
from meteoritic abundances and spectroscopic measurements
$Z/X=0.0245\pm 0.0015$ \cite{gr93}
, and finally
the age $(4.6\pm0.004)\, Gy$.
Furthermore the computation involve some additional hyphotesis 
and the use of an appropriate theory (e.g. mixing--length) for the treatment of the convection, to simplify 
the theoretical descriptions.

Here we will show results obtained by using two reference models -- Model S -- by Christensen--Dalsgaard et al.
 \cite{CD96}, which use respectively
 the OPAL \cite{ro96} equation of state,
and the MHD \cite{mi88} equation of state.
The MHD equation of state, based on the `chemical' picture of the plasma,
takes into account the effect of excited levels of atoms and ions on the properties of plasma and it also considers a lowest-order Coulomb coupling term 
through the Debye--H\"uckel approximation.
The OPAL equation of state, in contrast, is based on a `physical' description, in which
nuclei and electrons (free or bound) are the only fundamental constituents of the thermodynamic ensemble.

It is important to point out 
that much of the uncertainty in a solar model rely on the physics 
which describe the surface, since
there are substantial difficulties in
modelling convective motions
and the thermodynamic properties of this region
as well as in
the treatment of non-adiabatic effects on the oscillations.
In most cases, in fact, the frequencies are calculated in the adiabatic 
approximation, which is certainly inadequate in the near-surface region.

We will limit here the considerations on the solar modelling, since this
 is not subject of the present review. General presentations and more detailed theory about computation of standard solar models are described in a number of standard texts, e.g. \cite{sc58}, \cite{cl68}, \cite{co68}, 
\cite{ki91}.

\subsection{Forward analysis}

\begin{figure}[tb]
 \begin{center}
\includegraphics[width=.496\textwidth]{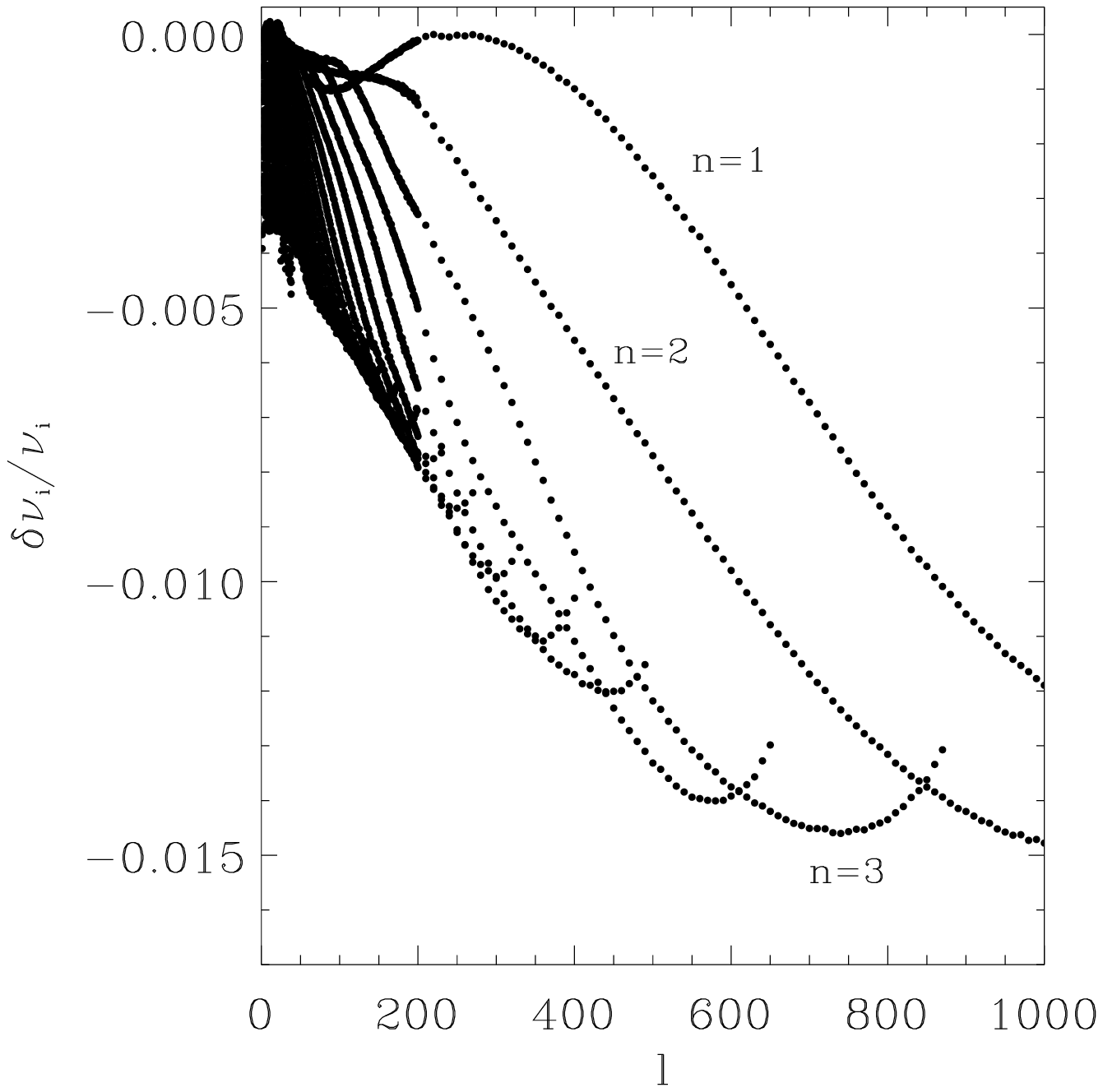}
\includegraphics[width=.496\textwidth]{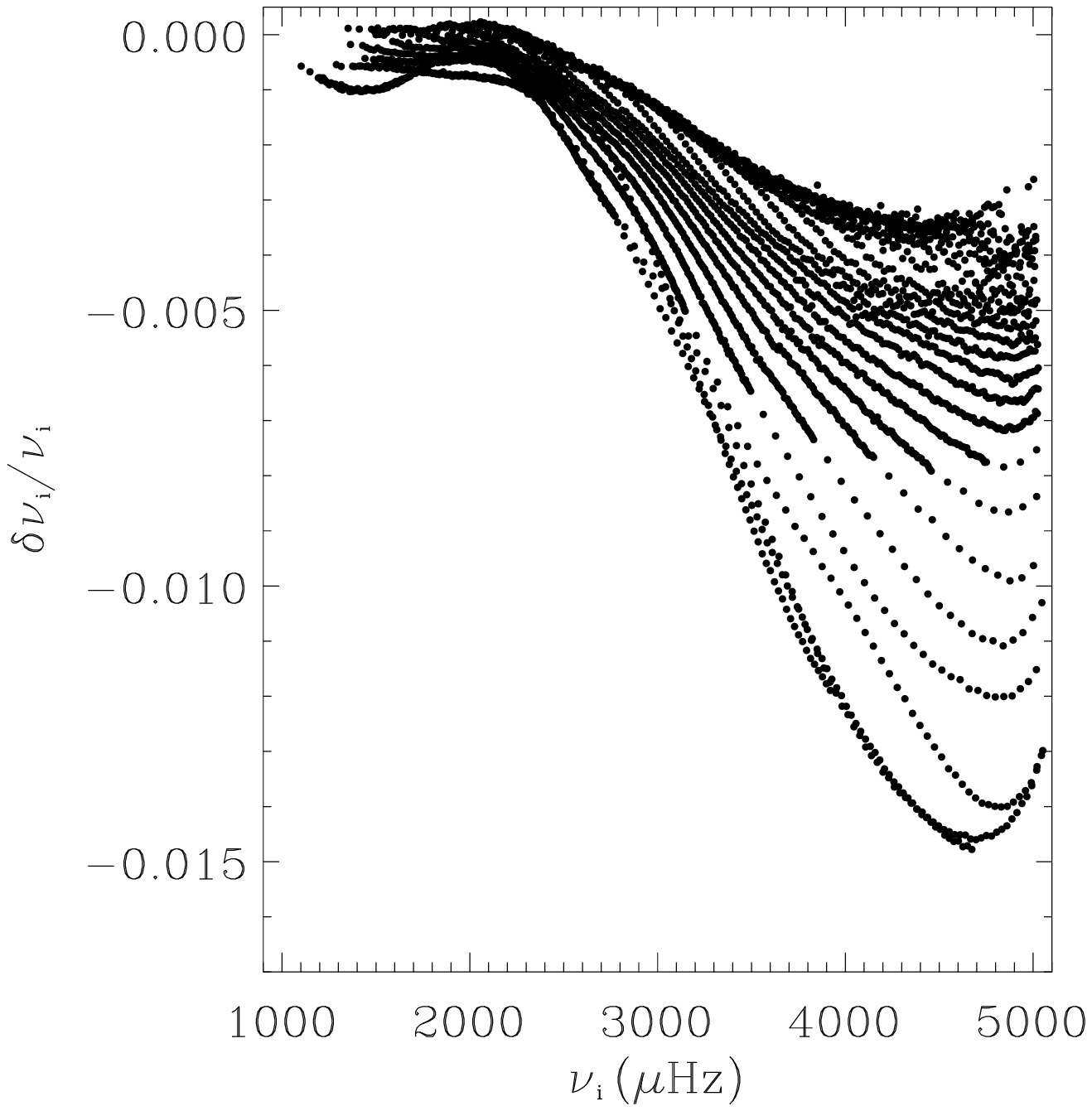}
\end{center}
\caption{Relative differences between the 
observed frequencies \cite{rh98} obtained by MDI instrument on SOHO satellite 
and the theoretical
frequencies computed on the reference model, as function of the degree 
(panel on the left) and of the frequency (panel on the right)}
\label{dif}
\end{figure}

A direct way to test a solar model is to consider differences between
observed frequencies and those calculated for the theoretical model.
The aim of this kind of investigation is to correct the physics on which 
is based the solar model in such a way to reduce the discrepancies.
Among several models, then, we should adopt the one which best fits the 
observed data.

Historically, one of the main successes of this approach was the spectacular
overall agreement of the theoretical $k_h-\omega$ diagram, produced by a standard model, with the observed one, showed in the 1988 by 
Libbrecht \cite{li88}.

Today we have the possibility to handle more accurate
 helioseismic observations.
Here, we discuss the results produced by considering 
helioseismic data \cite{rh98}
 obtained in 1996 by the MDI instrument on board the 
SOHO satellite \cite{sc95}.

Figure~\ref{dif} shows
 the relative differences between the observed frequencies
 and those calculated by using
Model~S by Christensen--Dalsgaard et al. \cite{CD96} as our
reference model and it employs the OPAL \cite{ro96} equation of state.
The differences between the observed and calculated eigenfrequencies
at low degree are very small
and vary slowly with the frequency.
 However, at high degree
 the differences appear to depend on $l$ and to increase with the frequency.

\begin{figure}[hb]
 \begin{center}
\includegraphics[width=.496\textwidth]{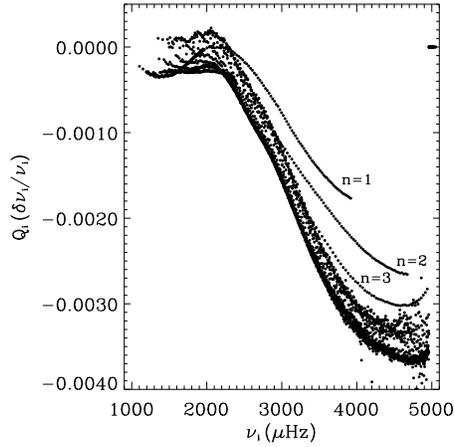}
\end{center}
\caption{Relative differences between the 
observed frequencies \cite{rh98} and the theoretical ones (like in Fig. 
\ref{dif})
 scaled by the inertia of the mode and plotted as function of frequency}
\label{Qdif}
\end{figure}

The frequency dependence results
in part from uncertainties in
the mode physics, but also from the real differences between the Sun and its
reference model.

The $l$-dependence is mainly associated with the variation of 
the mode inertia, since modes with higher $l$ penetrate less deeply and 
hence have a smaller inertia. Thus, high-degree modes are affected 
more strongly by the near-surface uncertainties \cite{an95} and 
\cite{go95}.
The $l$-dependence can be isolated in part  
 by considering frequency differences scaled  
by the inertia of the modes or more conveniently scaled by $Q_i$ that
 is the inertia of the mode $i=(n,\,l)$, normalized by the inertia of a
radial mode of the same frequency (see Fig. \ref{Qdif}).
The result is that the major 
inconsistencies, which appear at high frequency,
 derive from the modelling of the surface layers, indicating
 that the physics applied there, is inadequate for describing the 
relevant phenomena.

\subsection{Solar seismic radius}

The radius of the Sun can
 be simply determined from measurement of its distance
and of the angular diameter of the visible disk. But, in practice 
it is not an easy task to distinguish the observed radius from   
the photospheric radius, defined as being the depth
 where the temperature equals the effective temperature.

The measurements of the photospheric radius obtained during the past from 
several groups using different instruments have provided  
results, which appear fairly consistent with the standard value,
 quoted by Allen~\cite{al76}.
 
Only in the 1997, Schou et al. \cite{sc97} succeed for the first time, 
in obtaining
 an helioseismic determination
of the solar radius by using high-precision measurements of oscillation 
frequencies of the f modes of the Sun, obtained from the MDI experiment
 on board the SOHO spacecraft.  They determined that 
the seismic radius is about $300\, km$ smaller than the model radius.
 A similar conclusion was reached by
Antia \cite{an98} on the basis of analysis
of data from the GONG network.

The helioseismic investigation of the solar radius
 is based on the principle that
the frequencies of the f modes of intermediate angular degree depend primarily
on the gravity and on the variation of density in the region below the surface, where the modes propagate. From the asymptotic dispersion relation 
(\ref{fmodes}),
one can easily deduce that $\omega\propto R_{\odot}^{-3/2}$. Therefore,
 by applying a variational principle,
 we can obtain a relation between f-mode frequencies, 
$\omega_{l,0}=2\pi\nu_{l,0}$, and the correction $\Delta R$ 
that has to be imposed to the photospheric radius $R_{\odot}$ assumed for the 
standard solar model:
\begin{equation}
\frac{\Delta R}{ R_\odot}=-
\frac{2}{3}\left<{\Delta\nu_{l,0}\over\nu_{l,0}}\right>\; ,
\label{deltar}
\end{equation}
where $<>$ denotes the average weighted by the inverse square of the
measurement errors.

\begin{figure}[tb]
 \begin{center}
\includegraphics[width=.68\textwidth]{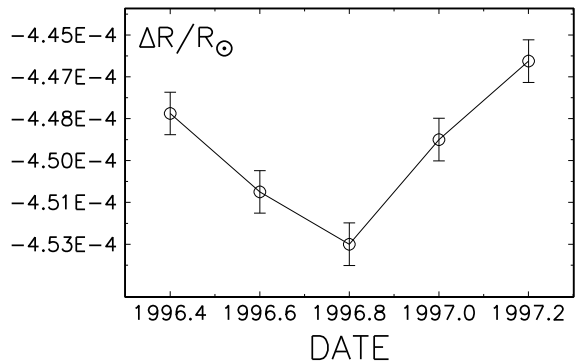}
\end{center}
\caption{Relative differences in radius of the Sun inferred from the 
f-mode frequencies and the standard solar model as in \cite{dz98}}
\label{radius}
\end{figure}

Dziembowski et al. \cite{dz98}, by analyzing a long time-series of 
MDI f-mode frequencies, 
 inferred temporal variation of the solar radius, with the aim 
of determining a possible solar cycle dependence.
 Their first results (Fig. \ref{radius}),
 covering the period from May 1996 to April 1997, show  that
the maximal relative variation of the solar radius during the observed period 
was about $\Delta R/R_{\odot}=6\times 10^{-6}$, which corresponds to approximately $\Delta R=4\, km$.
Recently, Dziembowski et al.~\cite{dz00} analyzing a larger set 
of data spanning a period from mid 1996 to mid 1999, have found that the
systematic trend of $\Delta R/R_{\odot}$ is not correlated with the magnetic activity.

However, Brown \& Christensen--Dalsgaard \cite{br98}, by combining
photoelectric measurements with models of the solar limb-darkening function,
determined
a photospheric radius of $695.508\pm0.026\; Mm$. 
This value appears
even smaller than the helioseismic one and the reason for this discrepancy 
is still unknown.    

It is clear that the 
problem that remains to be clarified is the connection between the 
 seismic radius determined from helioseismic measurements 
and the definitions of solar radius as obtained from the other methods.

\section{Helioseismic inversion}

The inverse problem, always associated with the forward approach, involves
estimating some functions to describe the physical properties of the Sun,
by solving integral equations
which appear expressed in terms of the experimental data.
Inversion techniques are well known and applied with success in several 
branches of the physics from geophysics 
to the radiation theory, as reported
 in \cite{pa77}, \cite{cr86} and \cite{ta87}.
Applications to the helioseismic data have been studied extensively 
during the last decade and
inversion methods and techniques have been
 reviewed and compared by several authors, e.g. 
 \cite{go91}, \cite{da91}, \cite{ko93},
\cite{dz94}, \cite{an94} and \cite{go96}. 

The observed
 data are related to the physics of the solar structure,
 in a very complicated way, and
the main difficulty arises from the fact that
 the helioseismic inversion is an
ill-posed problem:
\begin{itemize}
\item For each set of data, there exists an infinite number of solutions 
\item The solution is not unique
\item The solution does not depend continuously on the data. 
\end{itemize}
In fact, the observed frequencies constitute a
finite set of data and 
 the errors in the observations prevent the solution from being determined 
with certainty.
Thus, an appropriate choice of a
 suitable technique of inversion is
the first important strategy to adopt during a helioseismic inverse analysis.

The first attempts at inversion used analytical methods to solve
 integral equations obtained in first approximation
 by applying the asymptotic dispersion relation of solar frequencies, 
the so-called Duvall law \cite{du82}. This inversion method is 
considered however not very accurate, since the Duvall law
 represents a rough approximation for  
the low degree modes which are the more appropriate to study the Sun's core.
For this reason, here we will consider only numerical techniques of inversion.

\subsection{Inversion Techniques}

Since most of the fundamental aspects of inversion do not depend 
on the dimensions of the space in which the problem is posed, 
for simplicity here we consider the general case of the 
linear one-dimensional inversion problem,
in which
the measured data 
$d_i$ are functionals of a single function, $f(r)$, of 
the distance to the centre $r$:
\begin{equation}
d_{i}=\int_{0}^{R_{\odot}}{\cal K}^{i}(r)f(r)\D r+\varepsilon_{i}~ ~ ~ ~
i=1,\ldots M
\label{int}
\end{equation}
where
$R_{\odot}$ is the radius of the Sun. 
The properties of the inversion
 depend both on the mode selection $i\equiv(n,l)$
and on the observational errors $\varepsilon_i$, 
which characterize {\it the mode set}
($i= 1, \ldots, M$) to be inverted.

The observational errors $\varepsilon_i$ in the data,
are assumed to be independent and Gaussian-distributed
with zero mean and variances $\sigma_i^2$.
Given a set of data and errors, 
the problem is to determine $f(r)$ by solving the Eq. (\ref{int}), where
 ${\cal K}^{i}(r)$,
the kernels of the integral, are known functions which depend on the 
quantities of the reference model and its eigenfunctions.

There are two important classes of methods of obtaining estimates of $f(r)$: 
optimally localized averaging method based on the original idea of
Backus \& Gilbert \cite{ba68}, \cite{ba70} and 
regularized least-squares fitting method due to Phillips \cite{ph62} and
Tikhonov \cite{ti63}. Both methods give linear estimates of 
the function $f(r)$ and give results in 
general agreement, as it was demonstrated by Christensen--Dalsgaard et al. \cite{CD90} and by
Sekii \cite{se97}.

\subsubsection{Optimally localized averaging (OLA)}
The 
 localized averaging kernel method allows us to solve Eq. (\ref{int})
by estimating a
localized weighted average of the unknown generic quantity $f(r)$ 
at selected target radii $r_{0}$'s by means of a linear combination of 
all the data $d_{i}$:
\begin{equation}
\bar{f}(r_{0})=\sum_{i=1}^{M}\alpha_{i}(r_{0})d_{i}=\sum_{i=1}^{M} \alpha_{i}(r_{0}) \int_{0}^{R_{\odot}} {\cal K}^{i}(r) f(r) \D r \; ,
\label{bg1}
\end{equation}
where $\alpha_{i}(r_{0})$ are the inversion coefficients to be found and 
\begin{equation}
K(r_{0},r)=\sum_{i=1}^{M} \alpha_{i}(r_{0}){\cal K}^{i}(r)
\end{equation}
are the so called 'averaging kernels'.

Because of the ill-conditioned nature of the inversion problem, 
it is necessary to introduce a regularization procedure.
By varying a trade-off parameter $\theta$,
we look for the coefficients $\alpha_{i}(r_{0})$ which minimize the propagation of the errors
and the spread of the kernel:
\begin{equation}
 \int_{0}^{R_{\odot}}J(r_0,r)K(r_{0},r)
^2 \D r+\tan \theta \sum_{i=1}^{M}\sigma_{i}^2
\alpha_{i}^2(r_{0})\; ,
\label{ls}
\end{equation}
assuming that
\begin{equation}
\int_{0}^{R_{\odot}}
K(r_{0},r)\D r=1 \; .
\end{equation}
$J(r_0,r)$ is a weight function
that is small near $r_0$ and large elsewhere, assumed to be:
\begin{equation}
J(r_0,r)=(r-r_0)^2\;.
\end{equation}

From Eq. (\ref{ls}) we obtain the expression for
 the inversion coefficients:                                   
\begin{equation}
 \alpha_{i}(r_{0})=\sum_{j=1}^{M}[S_{i
j}(r_0)+\tan \theta E_{ij}]^{-1}\int_{0}^{R_{\odot}}{\cal K}^{j}(r)\D r\; ,
\label{coef}
\end{equation} 
where
\begin{equation}
S_{ij}(r_0)=\int_{0}^{R_{\odot}}(r-r_{0})^{2}
[{\cal K}^{i}(r){\cal K}^{j}(r)] \D r\; ,
\end{equation}
and the diagonal covariance matrix of the errors has elements:
\begin{equation}
E_{ij}=\left\{\begin{array}{l@{\quad}l}
\sigma_i^2\alpha_i^2(r_0) & \mathrm{for}\,\, i=j \\
0\,\,\,\,  & \mathrm{for}\,\, i \not= j 
\end{array} \right.
\end{equation}

By lowering the trade-off
 parameter it is possible to obtain more
localized averaging kernels closer to the nominal point 
$r=r_{0}$, but
 this decreases the accuracy with which the solution is determined, since the 
importance of the errors increases.
Thus, we should choose, among all the possible solutions, an optimal
compromise between localization and accuracy 
of the solution.

The Eq. (\ref{coef}) is equivalent to solve the following
set of linear equations:
\begin{equation}
 \mathrm{A}(r_{0})\vec{\alpha}(r_0)={\bf b}\; ,
\end{equation}
where $\vec{\alpha}(r_0)$ represents the vector for each target radius,
whose $M$ elements are
the coefficients
$\alpha_i(r_0)$;
${\bf b}$ is the vector which contains
the Lagrangian multipliers; $\mathrm{A}(r_0)$ is the $M \times M$ symmetric matrix, whose elements for each $r_0$ are $a_{ij}\!=\!S_{ij}(r_0)+\tan \theta E_{ij}$.
Therefore, 
the OLA method
is very much demanding on computational resource, since it requires
the inversion of $N$
 matrixes of order $M$ to determine the solution at $N$ radial points.

The errors of the solutions are the standard deviations calculated in the following way:
\begin{equation}
\delta \bar{f}(r_0)=\left[\sum_{i=1}^M \alpha_i^2(r_0)\sigma_i^2\right]^{1/2}\; ,
\label{deltaf}
\end{equation}
while the radial spatial resolution is assumed to be the half-width at half-maximum of the resolving kernels.

The same method can be applied in the variant form described by 
Pijpers and Thompson  in \cite{pi92} and \cite{pi94}, 
known as SOLA method (Subtractive Optimally Localized Averaging),
 making attempts to fit the averaging kernel to a target function, usually a 
Gaussian function 
$G(r_{0},r)$, of appropriate width and centered at the target radiulos:
\begin{equation}
K(r_{0},r)\simeq G(r_{0},r)\simeq\delta(r_{0},r)\; .
\end{equation}
 In this case, the trade-off parameter is rescaled at each target location to keep constant the width of the averaging kernels and to obtain more localized resolving kernels closer to the nominal concentration point. Therefore, the coefficients are determined by minimizing the following:
\begin{equation}
 \int_{0}^{R_{\odot}}
[\sum_{i=1}^{M}\alpha_{i}(r_{0}){\cal K}^{i}(r)-G(r_{0},r
)]^2 \D r+\tan \theta\sum_{i=1}^{M}\sigma_i^2
\alpha_{i}^2(r_{0})\; ,
\label{sola}
\end{equation}
so that
\begin{equation}
\alpha_{i}(r_{0})=\sum_{j=1}^{M}(U_{i
j}+\tan \theta E_{ij})^{-1}\int_{0}^{R_{\odot}}{\cal K}^{i}(r){G}(r_0,r)\D r\; ,
\end{equation} 
where
\begin{equation}
U_{ij}=\int_{0}^{R_{\odot}}
[{\cal K}^{i}(r){\cal K}^{j}(r)] \D r \; .
\end{equation}

This inversion problem appears equivalent to solve 
the following set of linear equations:
\begin{equation}
\mathrm{W}{\vec \alpha}(r_{0})= {\bf g}(r_{0})\; , 
\end{equation}
where $\mathrm{W}$ is a matrix
whose elements are
$w_{ij}=U_{ij}+\tan \theta E_{ij}$, 
 and hence does not depend on $r_0$; ${\vec g}(r_{0})$ is the cross-correlation vector of the kernels with the target function $G(r_0, r)$.
Thus, the solutions are obtained by inverting the matrix $\mathrm{W}$ only one time, such that the computational efforts is therefore reduced substantially.

\subsubsection{Regularized Least-Squares Fitting (RLS)}

This method allows to find a solution that is 
expressed as a linear combination of a chosen set of base functions 
$\phi_j$ with $j=(1,\ldots,N)$:
\begin{equation}
\bar{f}(r)=\sum_{j=1}^{N}f_{j}\phi_j(r)\; ,
\end{equation}
where $f_j$ are constants to be determined.

We can  choose the base functions, for example, being piecewise constant functions on a
dissection $r_j$ of the interval $[0,\,R_{\odot}]$:
\begin{equation}
\phi_j(r)=\left\{\begin{array}{r@{\quad }l}
1 & r_{j-1}<r<r_j \\ 
0 & \mathrm{elsewhere} \end{array} \right.
\end{equation}
so that $\bar{f}(r)=f_j$ on the interval $[r_{j-1},\, r_j]$.
Other common alternatives  
are to choose $\phi_j(r)$ as a continuous set of 
piecewise linear functions or as a set of splines.

The parameters $f_j$ are determined by a least-squares fit to the data.
However, this procedure needs a regularization procedure to obtain a 
smooth solution. So, basically we can determine the constants by minimizing:
\begin{equation}
\sum_{i=1}^M\frac{1}{\sigma_i^2}\left[d_i-\int_{0}^{R_{\odot}}{\cal K}^{i}(r)
\bar{f}(r)\D r\right]^2 +
\mu \int_{0}^{R_{\odot}}[{\cal F} \bar{f}(r)]^2\D r\; ,
\label{rls}
\end{equation}
where $\mu$ is a trade-off parameter between resolution and error 
 and ${\cal F}$ is a differential 
operator so that ${\cal F} \bar{f}(r)$ is a
suitable weight function that determines the relative importance of smoothing 
in different regions.

The minimization of the (\ref{rls}) leads to a set of linear equations 
which permits to determine $f_j$ and hence the solutions.

\section{Inversions for the solar structure}

\subsection{The variational principle}

The numerical inversion of data to determine
the solar structure is based on the use of
 the variational principle of Chandrasekhar \cite{ch64}.
Thus, the eigenfrequencies can be determined by solving an eigenvalue problem, whose expression can be obtained directly from the basic
 equations governing linear adiabatic oscillations Eqs.
(\ref{eqosc1})--(\ref{eqosc2}):
\begin{equation}
\omega^2\vec{\delta r}=\cal{F}(\vec{\delta r})\; ,
\label{var}
\end{equation}
where $\omega^2$ are the eigenvalues,
$\cal{F}$ is a linear operator on the eigenfunctions $\vec{\delta r}$.
 
Although the frequencies of solar oscillations can be known from observations, the eigenfunctions cannot be determined experimentally, so Eq. (\ref{var})
define a nonlinear integral equation.

However, Eq. (\ref{var}) can be
linearized around a known 
reference model, under the assumption of hydrostatic 
equilibrium. 
This procedure, whose details can be found, e.g., in \cite{un89},  
provides a linear integral equation 
 that can be used in an inverse procedure to determine the corrections which 
have to be imposed to the reference model 
in order to obtain the observed oscillation frequencies $\omega_i=2\pi\nu_i$.

\subsection{The surface term}
  
Non-adiabatic effects and other errors in modelling the surface layers, that 
can give 
rise to frequency shifts, as it was explained in Section 4.1, have to be taken into 
account by including an arbitrary function of frequency 
$F_{\rm surf}(\nu)$ in 
the variational formulation, as suggested by Dziembowski et al. in \cite{dz90}.

The function $F_{\rm surf}(\nu)$ must be determined as part of the analysis of the frequency differences. It should resemble,
 in practise, the differences $Q_{i}{\delta\nu_i}/{\nu_i}$ plotted in Fig. 
(\ref{Qdif}).
As $F_{\rm surf}(\nu)$ is assumed to be a slowly varying function of frequency, it can be expressed as expansion of Legendre polynomials  
$P_{\lambda}(\nu)$, usually of low degree $\lambda$.

In the inversion procedures
it is common use to suppress the surface term \cite{da91}.
This is done
 by constraining the inversion coefficients to satisfy:
\begin{equation}
\sum_{i=1}^{M} \alpha_{i}P_{\lambda}(\nu_{i})Q_i^{-1}=0\quad \quad \quad
\lambda=0,1\ldots \Lambda \; .
\label{fsurf} 
\end{equation}
The maximum value of the polynomial degree, $\Lambda$, used in the expansion is a free parameter of the inversion procedure. In practice, we should consider
 an appropriate value of $\Lambda$ for any given data set.

\subsection{Inversion for sound-speed and density}

It follows, from the preceding discussion, that
the differences in, for example, sound speed $c$ and density $\varrho$
between the structure of the Sun and the reference
model ($\delta c^2/c^2,\, \delta \varrho/\varrho$) can be expressed by the following integral equation \cite{dz90}:
\begin{equation}
\frac{\delta\nu_i}{\nu_i} =
\int_0^{R_{\odot}} K_{c^2}^i (r) \frac{\delta
c^2}{c^2}(r) \D r
+ \int_0^{R_{\odot}} K_{\varrho}^i (r) \frac{\delta \varrho}{\varrho}(r) \D r
 + \frac{F_{\rm surf}(\nu)}{Q_i} + \varepsilon_i  \; ,
\label{eqn:freqdif}
\end{equation}
where $K_{c^2,\varrho}^i$
and  $K_{\varrho,c^2}^i$ are
the kernels. The term $Q_i$ has been already introduced in Section 4.1.

Equation (\ref{eqn:freqdif}) 
forms the basis for the linearized structure inversion.
Unlike the case considered in Section 5.1,
this linearized inverse problem involves three unknown functions: $\delta
c^2/c^2$, ${\delta \varrho}/{\varrho}$ and $F_{\rm surf}(\nu)$.
However, the number of the unknown functions can be reduced to one
by adapting the method of the optimally localized averages.

The principle of the inversion, by generalizing the SOLA technique 
(see Eq. \ref{sola}), is to form a linear combinations of 
$\delta\nu_i/\nu_i$
with coefficients $\alpha_i(r_0)$ chosen to minimize:
\begin{equation}
\int_0^{R_{\odot}} \left[ {\cal K}(r_0,r) - {\cal G}(r_0,r) \right]^2 \D r +
\beta \int_0^{R_{\odot}}  {\cal C}^2 (r_0,r) \, f(r) \, \D r 
+ \mu\, \sum_{i=1}^{M} \alpha_i^2(r_0) \sigma_i^2 \; ,
\label{eqn:SOLA}
\end{equation}
where 
\begin{equation}
{\cal K} (r_0,r) = \sum_{i=1}^{M}\alpha_i(r_0) K_{c^2,\rho}^i(r) \; 
\label{eqn:avker}
\end{equation}
are the {\it averaging kernels}, 
while 
\begin{equation}
{\cal C} (r_0,r) = \sum_{i=1}^{M} \alpha_i(r_0) K_{\rho,c^2}^i(r) \; 
\label{eqn:crosst}
\end{equation}
are the {\it cross-term kernels}.
The parameter $\beta$ control the balance between
 the contribution from
$\delta \varrho/\varrho$ on $\delta c^2/c^2$;
 $\mu$ is the trade-off parameter,
 determining the balance between
the demands of well-localized kernels and a small error in the solution;
 $f(r)$ is a suitably increasing function of radius aimed at
suppressing the surface structure in the cross-term kernel, e.g. we can use
 $f(r) = (1 + r/R)^4$.

Thus, if our goal is to infer the speed of the sound, the 
coefficients $\alpha_i(r_0)$
should be 
 chosen such to suppress the contribution from the cross term,
 to localize the averaging kernel near $r=r_0$, to suppress the surface term 
 assuming the (\ref{fsurf}),
 while limiting the error in the solution, by the use of the two parameters
$\beta$ and $\mu$.

\subsubsection{Inversion results}
The first significant results concerning the application  of the inversion
 technique to the Sun were obtained in 1985 by Christensen--Dalsgaard et al.
\cite{CD85}, who produced the sound speed profile 
in the interior of the Sun and 
 who first determined the location of the base of the convection zone. 
\begin{figure}[b]
 \begin{center}
\includegraphics[width=.75\textwidth]{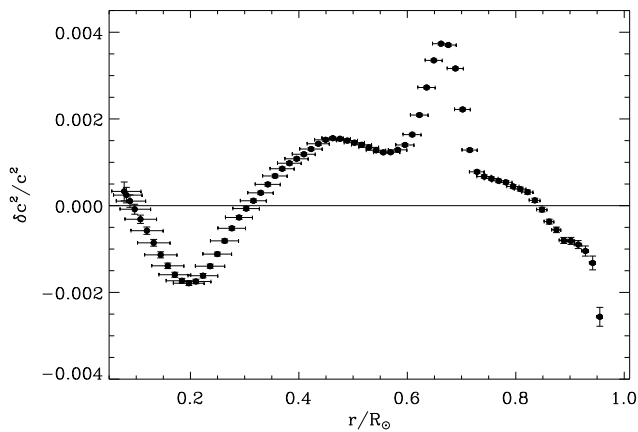}
\end{center}
\caption{The relative squared sound-speed difference between the Sun and the 
standard solar model \cite{CD96} as obtained by inversion of MDI/SOHO data \cite{sc98a}}
\label{Fc2}
\end{figure}
Since then, several efforts have been done for inverting data in order to test the correctness of the standard models in view of the improvements accomplished in the description of the relevant physics. A significant progress, in particular, has been achieved with the inclusion of diffusion of helium and heavy elements at the base of the convective zone \cite{ba96}.

 The resulting profiles for the speed of the sound and for the density, which are
 shown here, have been obtained by inversions of
 high quality helioseismic data 
obtained during 1998 by Schou \cite{sc98a},
 from SOI--MDI \cite{sc95}  instrument
 on SOHO satellite. This set includes only
modes with harmonic degree $l\leq 100$.
 
The Model~S \cite{CD96} which employs the OPAL \cite{ro96} equation of state
is used here as
reference model.

\begin{figure}[t]
\begin{center}
\includegraphics[width=.75\textwidth]{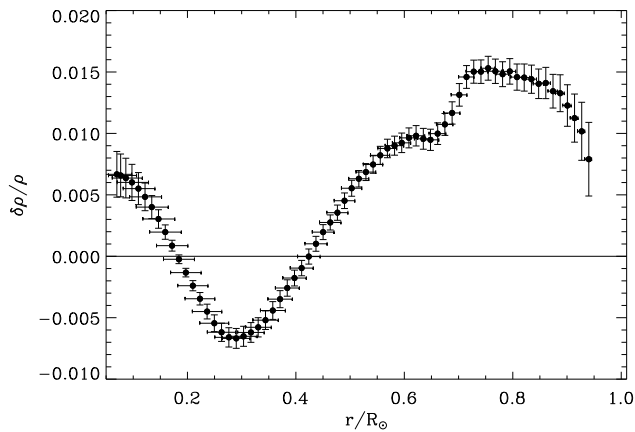}
\end{center}
\caption{The relative squared density difference between the Sun and standard
solar model \cite{CD96} as obtained by inversion of MDI/SOHO data \cite{sc98a}}
\label{rho}
\end{figure}

Figures \ref{Fc2} and \ref{rho} show the behaviour of the relative squared 
sound-speed and density differences between the Sun and the standard solar model as function of the fractional radius. 
The vertical error bars correspond to the standard deviations 
 based on the errors in the mode sets, calculated by Eq. (\ref{deltaf}),
whereas the horizontal bars give a measure of the localization 
of the solution.
The results indicate the substantial correctness of the standard solar models.
In fact,
it is clear that deviations are extremely small, except below the base of the convection zone ($0.71\,R_{\odot}$) where the theory fails to correctly describe the turbulent convection.

The structure of the core, however, is still quite uncertain since the few modes with lowest harmonic degree that are able to penetrate towards the centre,
 sample the core for a relative short time because of the large sound speeds there.

\subsection{Inversion for equation of state and the solar helium abundance}

The equation of state can be investigated  through the first
 adiabatic exponent $\Gamma_1$, the partial logarithmic derivative of pressure with respect to density at constant specific entropy,
already defined in Eq. (\ref{gamma}).

The solar plasma is almost an ideal gas,
 and the first adiabatic exponent is therefore close to $5/3$ in most 
of the interior.
It deviates from this value in the zones of hydrogen and helium ionization, near the surface.
Therefore, inversions of helioseismic data can be used, in particular,
to study the equation of state and to probe the helium abundance in the solar envelope, as it was 
proved e.g. in \cite{go84}, \cite{ko92}, \cite{dz92}, \cite{ba97}.

An integral equation analogous to Eq. (\ref{eqn:freqdif}) can be derived to 
determine the behaviour of $(\delta \Gamma_1/\Gamma_1)_{\rm int}$ the 
relative intrinsic difference in 
$\Gamma_1$, at constant pressure $p$, density $\varrho$ and composition, between the equation of state of the Sun and the one of 
the reference model, as in \cite{ba97}.

The kernels for $(c^2, \varrho)$ which appear in Eq. (\ref{eqn:freqdif})
can be converted
to kernels for the set
$(\Gamma_1, u, Y)$, where $u \equiv p/\varrho$ and $Y$ the helium abundance. 
After the conversion, Eq.~(\ref{eqn:freqdif}) can be written as
\begin{eqnarray}
\frac{\delta\nu_i}{\nu_i} =
\int_0^{R_{\odot}}  K_{c^2,\varrho}^i \left(\frac{\delta
\Gamma_1}{\Gamma_1} \right)_{\rm int} \D r
+ \int_0^{R_{\odot}}  K_{u,Y}^i \; \frac{\delta u}{u} \; \D r \nonumber \\
+ \int_0^{R_{\odot}}  K_{Y,u}^i \; \delta Y \; \D r
+ \frac{F_{\rm surf}(\nu)}{Q_i} 
+ \varepsilon_i  \; ,
\label{eqn:difg}
\end{eqnarray}
where $(\delta \Gamma_1/\Gamma_1)_{\rm int}$ is the difference in
$\Gamma_1$ that results from the differences
in the equation of state alone, but not from the resulting change
in solar structure. The term $\delta Y$ denotes the difference of
the helium abundance in the convective zone between the Sun
 and the model.

According to the
the OLA technique of inversion (Section 5.1), 
the coefficients are found by minimizing:
\begin{eqnarray}
\int_0^{R_{\odot}}  {\cal K}^2 (r_0,r) J(r_0,r) \D r +
\beta_1 \int_0^{R_{\odot}} \left( \sum_{i=1}^M \alpha_i(r_0) K_{u,Y}^i \right)^2 \, f(r) \, \D r \nonumber \\
+ \beta_2 \int_0^{R_{\odot}} \left( \sum_{i=1}^{M} \alpha_i(r_0) K_{Y,u}^i \right)^2 \, f(r) \, \D r
+ \mu\, \sum_{i=1}^M \alpha_i^2(r_0) \sigma_i^2 \; .
\label{eqn:MOLA}
\end{eqnarray}
\begin{figure}[t]
 \begin{center}
\includegraphics[width=.75\textwidth]{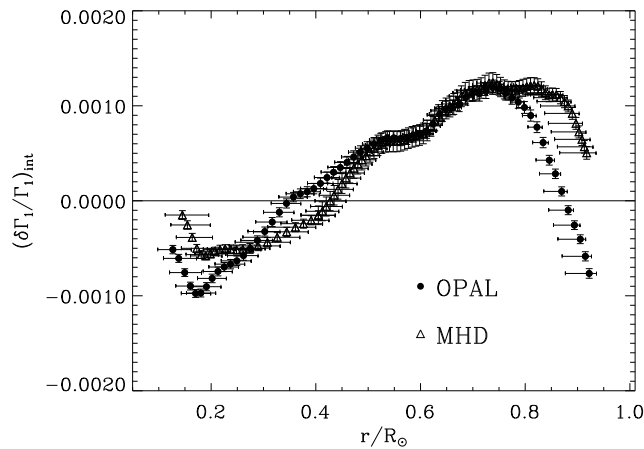}
\end{center}
\caption{The intrinsic difference in the adiabatic exponent 
$\Gamma_1$
 between the Sun and the OPAL \cite{ro96} equation of state (filled circle) and the Sun and the MHD equation of state \cite{CD96} (open triangles)
 obtained by inversion of a set of data by Schou \cite{sc98a}, which does not
include high-degree modes}
\label{gamma1ld}
 \end{figure}
The parameters $\beta_1$ and $\beta_2$ control the contributions of
$\delta u/u$ and $\delta Y$, respectively, and $\mu$ is a trade-off
parameter which controls the effect of data noise.
As in Eq.~(\ref{ls}) $J(r_0,r)$ is a weight function;
 $f(r)$ is included
to suppress surface structure in the 
first and second cross-term kernels, like in Eq.~(\ref{eqn:SOLA}).   

Figure \ref{gamma1ld} from Di Mauro \& Christensen--Dalsgaard \cite{di00a}, shows the resulting intrinsic differences in 
$\Gamma_1$ between the Sun and the two available equations of 
state (OPAL and MHD),
 as obtained by inversion of the data 
set by Schou \cite{sc98a}, which includes only modes with 
low and intermediate harmonic degree ($l\leq 100$).
 
As already shown by Basu \& Christensen--Dalsgaard \cite{ba97}, 
 by using only low and intermediate-degree modes it is difficult to judge
the significance of
 the differences between the two equations of state.
Nevertheless, Fig. \ref{gamma1ld} 
 confirms previous findings by
Elliott \& Kosovichev \cite{el98} that $\Gamma_1$ deviate from $5/3$ in the central core, probably due to relativistic effects.

\begin{figure}[t]
 \begin{center}
\includegraphics[width=.75\textwidth]{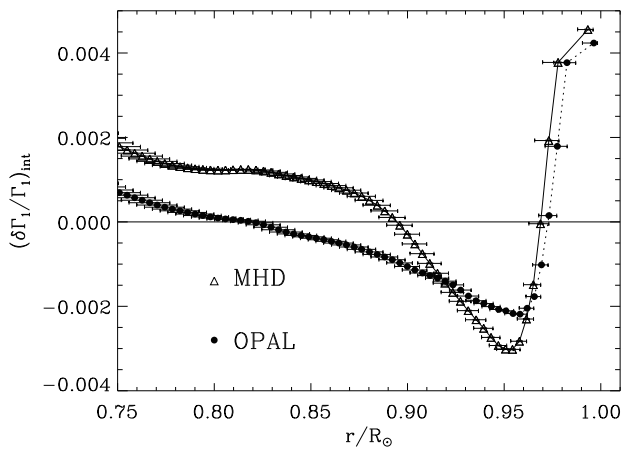}
\end{center}
\caption{The intrinsic difference in the adiabatic exponent 
$\Gamma_1$
 between the Sun and the OPAL \cite{ro96} equation of state
(filled circle) and the Sun and the MHD \cite{mi88} equation of state (open triangles)
 obtained by inversion of a set of data by Rhodes et al. \cite{rh98}, which includes high-degree modes}
\label{gamma1hd}
\end{figure}

The results shown in Fig. \ref{gamma1hd}, as found in Di Mauro \& Christensen--Dalsgaard \cite{di00b}, have been
carried
 out by inversion of preliminary
helioseismic data by Rhodes et al. \cite{rh98}, which include
high-degree modes ($l < 1000$),
 obtained in 1996 by the MDI instrument on board the 
SOHO satellite.
The set is made up of a
very large number of data
(7480 modes), which makes the 
computations slow and very demanding in terms of computer memory.
The precise high-degree modes  are able to determine variations
very near the solar surface, through the He~II ionization zone and also part of
the He~I ionization zone, while by using only low and intermediate-degree
 modes (Fig. \ref{gamma1ld}), we cannot determine solutions above $r \simeq 0.96 R_{\odot}$.

From Fig. \ref{gamma1hd} we can
 affirm that, as noticed by Basu et al. \cite{ba99}, the
 OPAL equation of state is able to describe better the plasma
 conditions in the interior of the Sun below $0.97\,R_{\odot}$.
In the upper layers above 
$0.97R_{\odot}$, the results indicate a large discrepancy 
between the models and the observed Sun, even considering
higher-order asymptotic terms in $F_{\rm surf}$. Here,
the use of very high degree modes reveals that 
 the differences between the 
two equations of state are very small.
  This is in contrast to earlier results by
 Basu et al. \cite{ba99}, which found evidence, by inverting a set of data
with no highest degree modes,
that MHD models give a more accurate description
of the very upper layers than the OPAL models.

\subsubsection{The helium abundance in the solar envelope}
It is well known that
spectroscopic measurements of the photospheric 
abundance of helium ($Y_{\odot}$) in the Sun are very uncertain and,
before the advent of helioseismology, the only accurate
method to quantify $Y_{\odot}$ was based on a calibration of solar models, in which the helium abundance has to be adjusted to match the observed solar luminosity.
The value of helium abundance calibrated with this method
 is typically about~$0.27$.

In the 1984, Gough \cite{go84} noted that the strong sensitivity of acoustic modes to the variation of the adiabatic exponent in the HeII ionization zone 
could allow also a seismic determination of the helium abundance in the outer layers of the Sun.
Thus, equations like (\ref{eqn:MOLA}),  may be inverted to determine $\delta Y$, the difference between the helium abundance of the Sun and that of 
the solar model in the helium ionization zones \cite{dz90}.
Since the convection zone is fully mixed, this provides a measure of 
the value of the helium abundance in the solar envelope.
It is also important to point out that the determination of the 
solar helium abundance, inferred from the inversion of data, is sensitive to the equation of state employed in the reference model.

The first seismic measures of $Y_{\odot}$ obtained by Christensen--Dalsgaard 
et al. \cite{CD96} reported values between $0.24$ and $0.25$, 
 that were significantly less than the abundance estimated
by the calibration on the standard solar model.
Dziembowski et al. \cite{dz91} pointed out that the difference was in rough agreement with that expected by the effect of gravitational settling of helium and heavy elements, as calculated by Cox et al. \cite{co89}. So, today settling is contained in all the most accurate standard solar models.

Recently, Di Mauro \& Christensen--Dalsgaard \cite{di00b} have used
 Eq. (\ref{eqn:MOLA}) to determine $\delta Y$, by inverting a set of data with
 high degree acoustic frequencies~\cite{rh98}.
By using the MHD equation of state, they
 obtained a value of $0.2426\pm0.0005$, 
consistent with the earlier results by Kosovichev \cite{ko97}
and Richard et al. \cite{ri96}, which employed
a similar variational technique.
By considering the OPAL equation of state they obtained  
a value of $0.2648\pm0.0004$,
which is strikingly higher than previous values quoted in the literature ($\simeq 0.242-0.25$)
by Basu \& Antia \cite{ba95}, Kosovichev \cite{ko97}, Richard et al. 
\cite{ri98} and Basu et al. \cite{ba99}.

The rather high value obtained for the helium abundance 
 based on the OPAL model
 may be due to the use of  
a set with very high degree modes.
The determination of observational frequencies for high-degree
modes still suffers from substantial difficulties, related to the
merging of power into ridges and the proper treatment of the leakage
matrix (e.g. \cite{sc98a}) and
this could cause systematic errors in the frequencies.

\section{Dynamics of the Sun}

\subsection{Fine structure in the acoustic spectrum of oscillations}
So far, we have considered only oscillations of a spherically symmetric 
structure, but it is well known and easily observed at the photosphere 
that the Sun is a slowly rotating star.

The rotation breaks the spherical symmetry of the solar structure and splits 
the frequency of each oscillation mode of harmonic degree $l$ into 
$2l+1$ components. Multiplets with a fixed $n$ and $l$ are said to exhibit a frequency ``splitting'' defined by:
\begin{equation}
\Delta \omega_{n,l,m}=\omega(n,l,m)-\omega(n,l,0)\; ,
\end{equation}
somewhat analogous to the Zeeman effect on the degenerate energy levels of an atom.

The determination of the splittings is often very difficult, so 
generally, the observations have not been applied in terms of individual 
mode frequencies, but rather it is customary to represent the frequency
splittings by polynomial expansion in terms of the so called 
$ a$-coefficients,
as explained in \cite{sc94}:
\begin{equation}
\omega(n,l,m)=\omega(n,l,0)+2\pi 
\sum_{j=0}^{j_{max}}a_{j}(n,l){\cal P}_{j}^{(l)}(m),
\label{split}
\end{equation}
where ${\cal P}_{j}^{(l)}(m)$ are orthogonal polynomials that can be chosen, for example, like Ritzwoller and Lavely in \cite{ri91}.
 Because of the symmetry properties of the splittings, the solar rotation is described only by the 
odd coefficients $a_{j}$, while
the even coefficients are a measure of the Sun's asphericity.

\subsection{Inversion for solar rotation}

To study the dynamics of the Sun we need to reconsider the derivation of the basic oscillation equations (\ref{eqmod1})--(\ref{eqmod3}) by including the effect of a velocity field. We assume that the rotation is sufficiently slow that the centrifugal force and other effects of second and higher order can be neglected.
This treatment allows
 to define a new expression which relates eigenfrequencies with
 eigenfunction and physical quantities, like the Eq. (\ref{var}).
By applying standard perturbation theory to the eigenfrequencies,
it can be shown that the rotational splittings 
are related to the rotation rate $\Omega(r,\theta)$ 
inside the Sun by:
\begin{equation}
\Delta \omega_{n,l,m}=
\int_{0}^{R_{\odot}}\int_{0}^{\pi}{\cal K}^{n,l,m}(r)\Omega(r,\theta)r 
\D r \D \theta 
\label{omega}
\end{equation}
where $\theta$ is the colatitude and ${\cal K}^{n,l,m}(r)$ 
are the mode kernel functions.
 The dependence of the splittings on angular velocity can be used in 
a 2-dimensional inverse problem to probe the dynamics of the Sun.

The 2-dimensional inverse problem can be simplified by considering that
the expansion of the splittings in polynomials given in the Eq. 
(\ref{split}) corresponds to an expansion of $\Omega(r,\theta)$ such that: 
\begin{equation}
\Omega(r,\mu)=\sum_{j=0}^{j_{max}}\widetilde{\Omega}_{2j+1}(r)\frac{\D P_{2j+1}(\mu)}{\D \mu}
\label{1.5}
\end{equation}
where $P_{2j+1}(\mu)$ are the Legendre polynomials with $\mu=\cos\theta$. 
So, the $a$-coefficients are related to the expansion functions  $\widetilde{\Omega}_{2j+1}(r)$ by:
\begin{equation}
2\pi a_{2j+1}(n,l)=\int_{0}^{R_{\odot}}{\cal K}
_{j}^{n,l}(r)\widetilde{\Omega}_{2j+1}(r) \D r 
\label{omegai}
\end{equation}
in which the kernels are calculated according to the expressions given
in \cite{di98b}.
  
Equation (\ref{omegai}) constitutes the basis for the 1.5-dimensional 
inversion.
Now, the original inverse problem Eq. (\ref{omega})
 has been decomposed into a series of 1-dimensional independent inversions for each $a$-coefficient to determine the expansion functions
 $ \widetilde{\Omega}_{2j+1}(r)$, whose combination according to Eq. 
(\ref{1.5}) leads to:
\begin{equation}
\Omega(r,\mu)= \widetilde\Omega_1(r)+\widetilde\Omega_3(r)\frac{dP_{3}(\mu)}{d\mu}+\widetilde\Omega_5(r)\frac{dP_{5}(\mu)}{d\mu}+......
\end{equation}

\begin{figure}[b]
 \begin{center}
\hspace{0.6cm}
\includegraphics[width=.6\textwidth,angle=270]{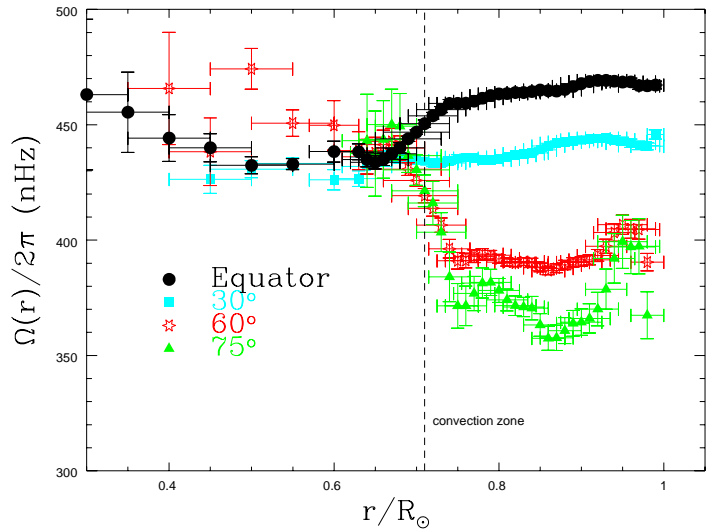}
\end{center}
\caption{Differential rotation at four latitudes as obtained by a 1.5 dimensional SOLA inversion of the SOI--MDI data. The approximate base of the convection zone is indicated by the dashed line \cite{di98b}}
\label{equat}
\end{figure}

\subsection{Inversion results}

The variation of the Sun's angular velocity with 
latitude and radius shown here,
 has been determined by Di Mauro et al. \cite{di98b}
 by means of a 1.5 dimension SOLA helioseismic inversion of more than 30,000
  $p$-mode splitting coefficients. These data were obtained from the 
first set of uninterrupted Doppler images from SOI--MDI (on board the 
SOHO satellite) in 1996 \cite{sc98b}, which yield splittings of 
great accuracy, never obtained in previous sets of data.  

The inferred rotation rate is shown in Fig. \ref{equat} where the points indicate the angular velocity at various depths calculated at the equator, and at latitudes of $30^{\circ}$, $60^{\circ}$ and $75^{\circ}$. 

\begin{figure}[tb]
 \begin{center}
\hspace{0.6cm}
\includegraphics[width=.75\textwidth]{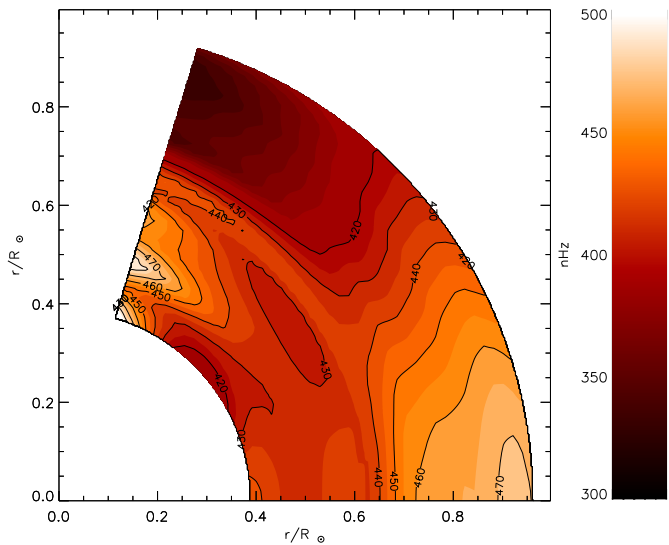}
\end{center}
\vspace{0.5cm}
\caption{Rotation rate in the Sun obtained by inversion of MDI data. Colours and contours indicate the isorotation surfaces. The white area indicates the region in the Sun where the data have no reliable determinations \cite{di98b}}
\label{sunred}
\end{figure}

In Fig. \ref{sunred} contours and red-scale indicate isorotation surfaces in a cut of the interior of the Sun.
 The results confirm the previous findings that the latitudinal differential rotation observed at the surface persists throughout the convection zone, while the radiative interior rotates almost rigidly at a rate of about $430\, nHz$. At low latitudes the angular velocity, through the
largest part of the convection zone, decreases with the radius while at high latitudes increases inwards. The near-surface behaviour agrees with the observed surface rotation rate.

The tachocline,
the transition layer from latitudinally-dependent rotation to nearly independent rotation \cite{sp92}, is of very considerable dynamical interest.
Furthermore, it is thought that the global dynamo behaviour, responsible for the solar 11 years
magnetic cycle, rises from strong toroidal magnetic fields generated by rotational shear in this thin region.

The tachocline appears
mostly located in the radiative zone at a pretty sharp midpoint near about $r=0.693\,R_{\odot}$ according to Corbard et al. \cite{co98}, and near $r=0.695\,R_{\odot}$ for Charbonneau et al. \cite{ch99}. It is also a fairly thin layer, not more than $0.05\, R_{\odot}$ at the equator. The layer seems 
to be wider at high latitudes, but certainly less than $0.1\, R_{\odot}$ \cite{di98a}. 
Charbonneau et al. \cite{ch99}, have recently confirmed that the width of the tachocline
appears to change with the latitude, with a minimum value at the equator
of $(0.0039\pm0.0013)\, R_{\odot}$.

Another interesting dynamical feature occurs near the poles, where unfortunately it is very difficult to localize the inversion solutions. 
Fig. \ref{equat} shows the presence at latitude of $75^{\circ}$ of a fairly localized region rotating faster than the surroundings \cite{sc98b}. It is still not clear, if this feature is somewhat related to the applied
 inversion technique.

Very recently, Howe et al. \cite{ho00} have 
 found evidence that the rotation rate near the base of the convective 
envelope shows variations with time,
with a period of the order of $1.3\,yr$ at low latitude. Such variations occur above and belove the tachocline and appear more pronounced near the equator and at high latitudes.

\begin{figure}[t]
 \begin{center}
\includegraphics[width=.55\textwidth, angle=270]{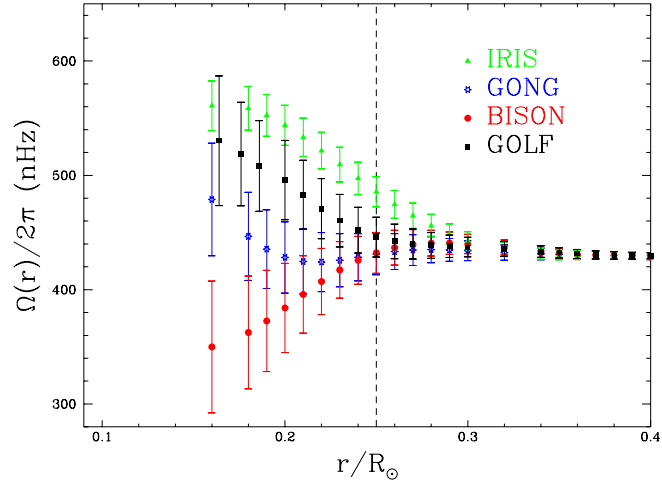}
\end{center}
\caption{Rotation of the Sun's core as deduced by inversion of the BISON (filled circles), IRIS (filled triangles), GONG (starred symbols) and GOLF (filled squares) sets of lowest degree splittings ($l=1-4$), all combined with MDI higher degree data set. The radial spatial resolution of each radial point is fixed at $\Delta r=0.1\,R_{\odot}$}
\label{rotcor}
\end{figure}

 To infer accurately the rotation in the deepest interior, it is necessary to invert a set of data which includes accurate splittings of the lowest degree modes ($l = 1-4$). 
The data sets, available for this purpose 
are obtained by the groundbased networks BiSON \cite{ch96}, IRIS \cite{la96} and GONG \cite{gav98} and from the GOLF \cite{ro98} instrument on SOHO.
 Unfortunately these sets of data 
 are not in mutual agreement and give conflicting 
 results of inversion in the core, as it is shown in Fig. \ref{rotcor}, taken from Di Mauro et al. \cite{di98b}.
Here, the radial spatial resolution, for clarity not drawn in the figure, is fixed at $\Delta r=0.1\,R_{\odot}$. 
  The independent sets of observations obtained by IRIS, GONG, GOLF lead to the conclusion that the Sun's core is in a state of rotation slightly faster than that observed at the surface in contradiction with the BiSON's data inversion which indicates a central angular velocity even slower than the surface polar angular velocity, as it was recently confirmed by Chaplin et al. \cite{cha99}.
Thus, the kinematics in the core remains largely uncertain, with a
 disagreement that might derive from the different data analysis procedures employed.

\subsection{Helioseismic determination of the solar angular momentum and quadrupole moment}

The present angular momentum of the Sun $\Im$, can be
 deduced from the internal rotational
behaviour derived from helioseismological data, 
 by integrating the following \cite{di98b}:
\begin{equation}
{\Im}=\int_{0}^{M_{\odot}}r^{2}\D M_r
\int_{0}^{1}(1-\mu)\Omega(r,\mu)\D \mu=\frac{2}{3}
\int_{0}^{M_{\odot}}\widetilde\Omega_1(r)r^2 \D M_r  \;,
\label{J}
\end{equation}
where $\widetilde\Omega_1(r)$ is determined by helioseismic inversion of the $a_1$-splitting coefficient, from Eq. (\ref{omegai}).
If we assume the angular velocity behaviour 
 shown in Fig. \ref{equat},
the integration of Eq. (\ref{J}) leads to 
$\Im=(1.96\pm 0.0
5)\cdot 10^{48}\, g\,cm^{2}\,sec^{-1}$ \cite{di98b}.
This value is in agreement within errors with the one obtained by Pijpers in \cite{pi98}.

Another quantity of particular interest is the gravitational quadrupole moment 
$J_2$ of the Sun, which can be deduced, according to Pijpers \cite{pi98}
 by evaluating the two-dimensional integral:
\begin{equation}
J_2=
\int_{0}^{R_{\odot}}\D r\int_{-1}^{-1}{\cal F}(r, \mu)\Omega^2(r,\mu) \D \mu\;,
\end{equation}
where ${\cal F}(r, \mu) $ is the two-dimensional kernel which depends
 on the physical quantities of the reference model and on some more general 
assumptions on the physics of the Sun.

The value of $J_2$
obtained by Pijpers \cite{pi98} is $J_2= (2.23\pm0.09)\times10^{-7}$.
This result is totally 
consistent with the one obtained by
Patern\`o et al. \cite{pa96} with a different approach based
on both
the measurement of solar oblateness and the angular velocity
 profile deduced by inversion of splittings.

\section{Seismology of the fine structure: solar asphericities}

The asymmetric part of the fine structure in the p-mode spectrum 
(Eq. \ref{split})
of solar 
oscillations varies in a systematic way through the solar cycle
\cite{ku88}, \cite{ku89},~\cite{li90}.

  It is evident that
the changes are associated with the surface temperature bands reported
by Kuhn et al. \cite{ku96}.  Also, Woodard \& Libbrecht \cite{wo91}
found a strong correlation between oscillation frequency changes and
solar surface magnetic variations from monthly averages of their data.
The origin of this behaviour, as
well as the temporal variation of the frequencies is still ambiguous, but
it appears clear that all these changes 
are consistent with a near surface perturbation.

The even-order splitting coefficients $a_{2k,l,n}$, seen in Eq. (\ref{split}) can be  
fitted to the following formula obtained by Dziembowski \& Goode \cite{dz91a}:
\begin{equation}
a_{2k,l,n}=a_{2k,l,n;{\rm rot}}+C_{k,l}{\gamma_k\over I_{l,n}},
\label{asph}
\end{equation}
where $a_{2k,l,n;{\rm rot}}$ represents the effect of 
centrifugal distortion which can be calculated following the treatment 
of Dziembowski \&  Goode \cite{dz92a}; $I_{l,n}$ is a measure of the 
modal inertia; $C_{k,l}$ is a constant which depends on the degree and on $k$, 
and $\gamma_k$ is the asphericity coefficients 
which is directly related to the distortion described by
the $P_{2k}(\mu)$ Legendre polymonial.
 The $P_2(\mu)$ term corresponds to a quadrupolar distortion (the oblateness), while $P_4(\mu)$ is the hexadecapole shape term and so on. 

\begin{figure}[bt]
 \begin{center}
\includegraphics[width=.6\textwidth]{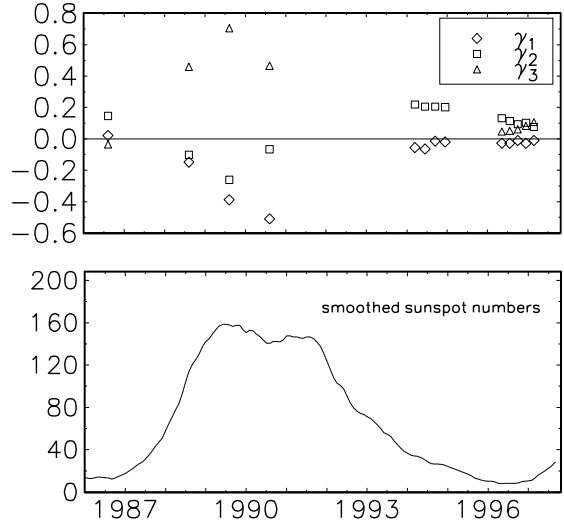}
\end{center}
\caption{ The lower panel is a smoothed monthly average of the 
sunspot number covering the time since the 1986 activity minimum.  
The upper panel is a combination of $\gamma$'s derived from 
BBSO (1986--90), LOWL (1994) and SOHO/MDI (1996--97) data.  The 
errors in the $\gamma$'s are smaller than the symbols used
to represent their values \cite{dz98}}
\label{aspher}
\end{figure}

Dziembowski et al. in \cite{dz98} have studied
the behaviour of the even splitting coefficients $a_{2k,l,n}$ of p modes,
 obtained by observations
covering almost all the period
 during the past 11 years cycle.
In Fig. \ref{aspher}, taken from \cite{dz98}, there are shown
 the mean values of the $\gamma_k$ coefficients as obtained by
 observations  
from various instruments, 
which include BBSO for the period 1986--1990, LOWL for the year 1994 and SOHO/MDI for the period 1996--1997. The variation of the asphericity coefficients
is compared in Fig. \ref{aspher} 
with the monthly averages of smoothed 
sunspot numbers. 
Clearly, the BBSO data of 1988 and 1989 give the
largest magnitudes of $\gamma_1$, $\gamma_2$ and $\gamma_3$, 
and this corresponds to the first half 
of the previous sunspot maximum.
In years of high activity 
 all three coefficients are substantial and change rapidly, while during
 low magnetic activity their value is roughly zero. 
In particular the asphericity appears more pronounced
 in period of high activity when it happens that 
$P_2(\mu)$ and $P_4 (\mu)$ distortions decrease while
 the $P_6(\mu)$ distortion increases. 
This can be translated in the fact that the Sun assumes a shape which varies
from simply oblate to complicated
asphericity according to the magnetic cycle.
  
This interesting conclusion has been confirmed by Howe et al. \cite{ho99}, which analyzed data obtained by the GONG network during the period 1995--1998. They also observed that 
 the temporal variation of the $a_{2k}$-coefficients is strongly
correlated with the latitudinal distribution of the surface
magnetic activity.

The behaviour of the $\gamma $'s as function of the frequency \cite{dz98} 
yields, also, information about
 the sources of the solar distortion reflected in the even-$a$ coefficients.
  It is well-known 
that p modes sample the region just above their inner turning
points. Recently, Dziembowski et al. \cite{dz00} found
 a significant
 aspherical distortion in the layer located at a depth ranging between 25 and 100 Mm. The perturbations seems to arise from a relative temperature increase 
of about $1.2 \times 10^{-4}$ or from a magnetic perturbation, with $\langle
B^2\rangle\simeq (60 \, KG)^2$.

\section{Concluding remarks}

Helioseismology, through the very accurate identification of
 oscillation frequencies of acoustic and fundamental modes, has clearly 
demonstrated that the standard solar models reproduce the behaviour of the Sun with remarkably accuracy, consistent within $1\,\%$.

 Despite such overall success, this discipline has not yet
exhausted its resources, since 
helioseismic results clearly suggest further refinements of the solar models.

The detailed structure of the convective zone and of the 
near-surface region is quite uncertain, since
there remains substantial ambiguity associated with modelling 
the convective flux, taking into account the
non-adiabatic effects, explaining the excitation and damping
of the solar oscillations 
 and defining an appropriate
 equation of state to describe the thermodynamic properties 
of the solar structure.

The attempts to restore the solar core conditions, up
to now, have been contradictory too.
In fact p modes (as opposed to gravity modes,
g modes) are not very sensitive to the core of the Sun. 
 This indicates the necessity of using more accurate low degree p-mode
data and to continue to investigate for the presence of g modes.

In addition, there is still much work ahead in getting a detailed 
understanding of the Sun's rotation. Some rotational 
features like, for example, the temporal changes which occur
 near the base of the convective envelope have not been yet explained.

Finally, by studying the connection between the seismic and the global characteristics of the Sun, the
challenge is to find the reason for the
correlation between the variation of the Sun's shape
and the magnetic solar cycle.

Ever more precise helioseismic observations from ground and space can help us
to reconstruct the complete picture of the Sun and, finally, to solve
 the most discussed open questions in solar physics such as the
 solar neutrino problem, the history of the Sun's angular momentum, and
the solar cycle generation mechanism, through the interaction 
of the convective motions with the rotation inside the Sun.

Recently, 
 a new window has been opened 
on the astrophysics research: the possibility to study
and to understand the behaviour of other stars by applying the tools 
and the techniques well developed and used in helioseismology.
In fact, the success of helioseismology has spurred
investigators to extend this diagnostic to other stars
 which may show multi-mode pulsations.
Up to now, the 
seismological study of pulsating stars, known as {\it Asteroseismology},
has been 
hindered by the problem of mode identification since the oscillation amplitudes observed on 
the Sun (a few parts per million in flux) are too small to be detected in other stars with
ground-based telescopes. To reach the required sensitivity and frequency resolution, several 
space experiments, MONS \cite{kj00}, COROT \cite{rox98},
MOST \cite{ma98}, will soon be devoted to the measurements of stellar oscillations. 
Thus, it is evident that asteroseismology represents the successive step in the
evolution of the helioseismology research.

\section*{Acknowledgements}
I thank the organizers of the {\it 5eme Ecole d'Astrophysique solaire}, particularly J.P.~Rozelot for an excellent and very enjoyable meeting.
The work presented here was supported in part by
the Danish National Research Foundation through its establishment
of the Theoretical Astrophysics Center and in addition by the Formation Permanente du CNRS (France).
I am very grateful to J. Christensen--Dalsgaard for his reading of an early version of this paper and for fruitful discussions and comments which have substantially improved this presentation.

\end{document}